\documentclass{aa}  

\usepackage{graphicx}
\usepackage{txfonts}
\usepackage{color}

\begin{document}

   \title{Dense cores and star formation in the giant molecular cloud Vela~C\thanks{Based on
         observations made with APEX telescope in Llano de Chajnantor (Chile) under ESO
         programme ID 089.C-0744 and MPIfR programme ID 087.F-0043}\fnmsep\thanks{Table~1 and 
         Table~2 are only available in electronic
          form}\fnmsep\thanks{Herschel is an ESA space observatory with science
          instruments provided by European-led Principal Investigator consortia and with important
          participation from NASA.}}


   \author{F. Massi
          \inst{1}
          \and
          A. Weiss\inst{2}
          \and 
          D. Elia\inst{3}
          \and 
          T. Csengeri\inst{2}
          \and 
        E. Schisano\inst{3,4}
          \and 
          T. Giannini\inst{5}
          \and 
        T. Hill\inst{6}
          \and
        D. Lorenzetti\inst{5}
          \and
          K. Menten\inst{2}
          \and 
          L. Olmi\inst{1}
          \and 
         F. Schuller\inst{7}
          \and 
         F. Strafella\inst{8}
          \and 
         M. De Luca\inst{9}
          \and 
         F. Motte\inst{10}
          \and 
         F. Wyrowski\inst{2}
          }

   \institute{INAF - Osservatorio Astrofisico di Arcetri,
              Largo E. Fermi 5, I--50125 Firenze, Italy
              \email{fmassi@arcetri.astro.it}
         \and
          Max-Planck-Institut f\"{u}r Radioastronomie,
          Auf dem H\"{u}gel 69, 53121 Bonn, Germany
          \email{csengeri@mpifr-bonn.mpg.de}
         \and
        INAF - Istituto di Astrofisica e Planetologia Spaziali, 
        Via Fosso del Cavaliere 100, I--00133 Roma, Italy
           \email{davide.elia@iaps.inaf.it}
         \and
        INAF - Istituto di Radioastronomia, and Italian ALMA Regional Centre, 
        via P. Gobetti 101, I--40129, Bologna, Italy
         \and
        INAF - Osservatorio Astronomico di Roma, via Frascati 33, 
        I--00040 Monte Porzio Catone, Italy
         \and
        Atacama Large Millimeter/Submillimeter Array, Joint ALMA Observatory, 
        Alonso de C\'{o}rdova 3107, Vitacura 763--0355, Santiago, Chile
         \and
        AIM, CEA, CNRS, Universit\'{e} Paris-Saclay, 
        Universit\'{e} Paris Diderot, Sorbonne Paris Cit\'{e}, 
        91191 Gif-sur-Yvette, France
         \and
         Dipartimento di Matematica e Fisica ‘Ennio De Giorgi’, 
         Universit\`{a} del Salento, CP 193, I--73100 Lecce, Italy
         \and
         \'{E}cole Normale Superi\'{e}ure, CNRS, Observatoire de Paris, 
         UMR 8112, LERMA, Paris, France
         \and
        Universit\'{e}  Grenoble Alpes, CNRS, IPAG, 38000 Grenoble, France
             }

   \date{Received ; accepted }

 
  \abstract
   {The Vela Molecular Ridge is one of the nearest (700 pc) giant molecular cloud (GMC)
    complexes hosting intermediate-mass (up to early B, late O stars) star formation, and is 
    located in the outer Galaxy, inside the Galactic plane.
    Vela C is one of the GMCs making up the Vela Molecular Ridge, 
    and exhibits both sub-regions of robust and sub-regions of more quiescent star formation activity,
    with both low- and intermediate(high)-mass star formation in progress. 
   }
   {We aim to study the individual and global properties of dense dust cores in Vela C, and aim to
    search for spatial variations in these properties which could be related
    to different environmental properties and/or evolutionary stages in the various sub-regions of 
    Vela C. 
   }
   {We mapped the submillimetre (345 GHz) emission from vela C with LABOCA 
    (beam size $\sim 19\farcs2$, spatial resolution $\sim 0.07$ pc
    at 700 pc) at the APEX telescope.
    We used the clump-finding algorithm CuTEx to identify the compact submillimetre sources. We also
    used SIMBA (250 GHz) observations, and {\it Herschel} and WISE ancillary data.
    The association with WISE red sources allowed the
    protostellar and starless cores to be separated, whereas the {\it Herschel} dataset allowed the
    dust temperature to be derived for a fraction of cores.
    The protostellar and starless core mass functions (CMFs) were constructed following two
    different approaches, achieving a mass completeness limit of $3.7$ $M_{\sun}$.
   }
   {We retrieved 549 submillimetre cores, 316 of which are starless and mostly
   gravitationally bound (therefore prestellar in nature).
   Both the protostellar and the starless CMFs are consistent with the
   shape of a Salpeter initial mass function in the high-mass part of the distribution. 
   Clustering of cores at scales of 1--6 pc is also found,
   hinting at fractionation of magnetised, turbulent gas. 
   }
   {
   }


\titlerunning{Dense cores and star formation in Vela~C}
\authorrunning{F. Massi et al.}

   \maketitle
%

\section{Introduction}
It has been known for a long time that the interstellar medium is arranged into a hierarchical distribution
of structures, from the most massive giant molecular cloud (GMC) complexes down to smaller
and less massive entities (namely clouds, clumps, and cores). 
At the smallest end of the
distribution lie dense cores, $\sol 0.1$ pc in size with densities of at least $10^{4} - 10^{6}$
cm$^{-3}$ (see, e.g. Bergin \& Tafalla \cite{be:ta}, Enoch et al.\ \cite{enoch07},
Andr\'{e} et al.\ \cite{andrewpp}). 
In particular, gravitationally bound starless cores are of the utmost importance, as they are
believed to be prestellar in nature and therefore the seeds of stars. In a seminal work, Alves et al.\ 
\cite{alves} used near-infrared (NIR)
extinction to derive the core mass function (CMF) in the \object{Pipe Nebula} ($d \sim 130$ pc)
with a spatial resolution of $0.03$ pc. These latter authors found that the shape of the CMF is remarkably
similar to the stellar initial mass function (IMF; e.g. Kroupa et al.\ \cite{kroupa},
Scalo \cite{scalo}, Chabrier \cite{chabry})
but scaled to higher masses by a factor of about three.
This is considered to be evidence that the stellar IMF is an outcome of the CMF, provided that an efficiency 
of $\sim 30$ \% results from the star-formation processes.

The availability of large bolometer arrays in the submillimetre (submm) and more recently of balloon-borne
(the Balloon-borne Large Aperture Submillimeter Telescope, BLAST) and space-borne (the {\it Herschel}
space observatory) far-infrared (FIR) telescopes has allowed the
CMF of prestellar cores 
in the nearest star-forming regions up to Orion to be sampled with high sensitivity and spatial resolution
(for a review, see Andr\'{e} et al.\ \cite{andrewpp}). These observations have confirmed that
the shape of the CMF is consistent with that of the IMF, strengthening the idea 
of a link between them. It is therefore fundamental to derive CMF and other prestellar core 
properties in regions with different levels and global efficiency of star formation to check whether  
the connection between IMF and CMF persists in different environments. In fact, significant differences
have recently been found in CMFs derived from observations of farther massive star-forming regions
(e.g. Motte et al.\ \cite{motte17}), now feasible thanks to the large mm interferometers. 
On the other hand, Olmi et al.\ \cite{olmi} used the Hi-Gal catalogue to study the {\it clump} mass 
function in several regions at a range of distances, finding again that its shape generally resembles
that of the IMF and suggesting that this supports gravito-turbulent fragmentation of molecular
clouds occurring in a top-down cascade.

The \object{Vela Molecular Ridge} (VMR) is a GMC complex first mapped in the CO(1--0)
transition (at low spatial resolution) by Murphy \& May \cite{MM}, who
divided the structure into
four main clouds (named A, B, C and D) corresponding to local CO(1--0) peaks. The VMR
is located in the
outer Galaxy, inside the Galactic plane, with a kinematic distance of 1--2 kpc.
Liseau et al.\ \cite{liseau} analysed all the distance indicators available, concluding that clouds 
A, C, and D are at $d= 700 \pm 200$ pc. 
In this regard, we performed a distance test using Gaia DR2 data (Gaia collaboration 2016, 2018), 
which is described in the online Appendix
(Appendix~\ref{sec:dist}). We found indications of a value of 950 pc (with an estimated error of
$\pm 50$ pc). Since this does not significantly affect our main
results, pending further refinements of the distance determination, we will adopt the value
of 700 pc for consistency with all previous works. However, we detail the effects of
the larger value in the Appendix. 

Subsequent observational studies have 
demonstrated that the VMR hosts low- and intermediate-mass (up to early B, late O stars) star
formation in a number of stellar clusters (Liseau et al.\ \cite{liseau},
Lorenzetti et al.\ \cite{loren}, Massi et al.\ \cite{mas1}, \cite{mas2}, \cite{massi03}).
It is therefore one of the nearest regions with intermediate-mass star formation in progress, just slightly
more distant than Orion. However, unlike Orion, the VMR is located in the Galactic plane, and is therefore
probably more representative of typical Galactic molecular cloud complexes. In this respect,
sampling the CMF in the VMR is valuable, since this region makes up a different environment from
those already studied in the solar neighbourhood.
 
\begin{figure*}
   \centering
   \includegraphics[angle=-90,width=12cm]{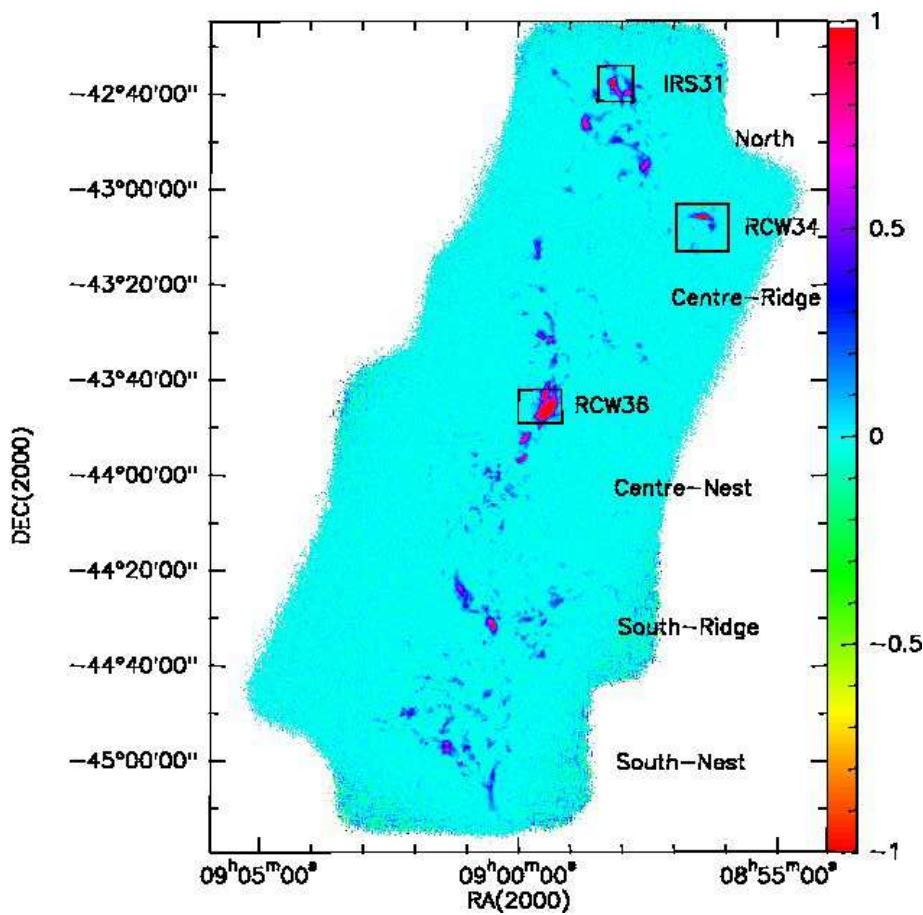}
   \caption{LABOCA map of 345 GHz emission towards Vela C. The colour scale
     is indicated in the bar on the right in units of Jy/beam. The locations of the 
     HII regions RCW~34 and RCW~36, and the young cluster IRS~31
     are enclosed with boxes and labelled (see Figs.~\protect\ref{rcw34:map}, 
     \protect\ref{rcw36:map}, and \protect\ref{irs31:map}, respectively, for 
     zoomed-in maps of these regions). The sub-regions defined by Hill et al.\ \cite{hill11} 
     are also labelled.
                      }
              \label{final:map}%
    \end{figure*}

The first large mm continuum mapping of the VMR was carried out by
Massi et al.\ \cite{massi07} towards \object{Vela D} at 250 GHz with SIMBA at the SEST telescope,
with a spatial resolution of $\sim 0.08$ pc, a little less than the
size of the largest dense cores. 
The VMR was then mapped in the FIR by Netterfield et al.\ \cite{Nett09} with
BLAST.  Subsequently, Olmi et al.\ \cite{holmes} combined the data
from Massi et al.\ \cite{massi07} and from Netterfield et al.\ \cite{Nett09}, obtaining
a CMF for Vela D whose shape is in good agreement with that of a standard IMF. 
Hill et al.\ \cite{hill11} studied \object{Vela C} using {\it Herschel} PACS and
SPIRE FIR/submm observations at 70, 160, 250, 350, and 500 $\mu$m, finding a filamentary
structure. They also identified five main sub-regions (see Fig.~\ref{final:map}) exhibiting
different morphologies (ridges and nests of filaments). Giannini et al.\ \cite{gianni12}
used the same data to retrieve dense cores and derive a CMF. They identified starless
cores based on the lack of emission at 70 $\mu$m and obtained a prestellar CMF
whose high-mass-end is flatter but still consistent with a standard IMF.

The {\it Herschel} PACS and SPIRE bands span the emission peak of cold dust, making these data
very useful for deriving the dust temperature. On the other hand, mm and submm continuum emission
from dust is more suitable for deriving column densities (and hence masses), since this occurs
in the optically thin regime. In addition, ground-based mm observations allow a better
angular resolution than space-borne instruments. 
Therefore, a better way to sample the CMF is by combining FIR
(dust temperature) and submm (column density) maps with comparable spatial resolution.
In this paper, we report on submm (345 GHz) observations of Vela C obtained with the
Large Apex Bolometer Camera (LABOCA) at the Atacama Pathfinder EXperiment
(APEX) telescope (see Fig.~\ref{final:map} for the final LABOCA map), 
with a spatial resolution of $\sim 19\arcsec$ ($\sim 0.07$ pc
at 700 pc). By combining ancillary
{\it Herschel} and Wide-field Infrared Survey Explorer (WISE) data, 
we have revised the CMF of Vela C obtained by 
Giannini et al.\ \cite{gianni12}. After outlining some well-studied specific regions of Vela C
in Sect.~\ref{vela:su}, 
we describe our observations, the sets of ancillary data, and the 
clump-finding algorithms used in Sect.~\ref{obsec}. Our results are presented in Sect.~\ref{resec}
and discussed in Sect.~\ref{dissec}. Finally, we summarise our work in Sect.~\ref{consec}.

\section{Young embedded star clusters and H{\sc ii} regions towards Vela C}
\label{vela:su}

Several individual star--forming regions in the VMR have been studied in detail by a 
number of authors. Three of these regions are relevant for Vela C and are described in the 
following.

\subsubsection{RCW~34}

\begin{figure}
   \centering
   \includegraphics[angle=-90,width=8cm]{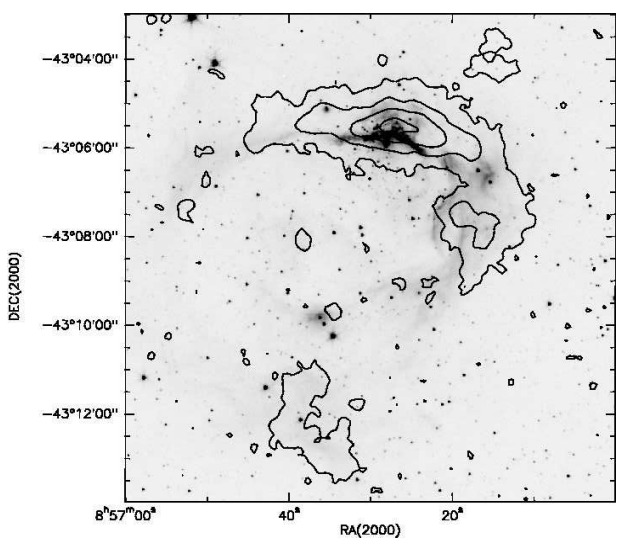}
   \caption{IRAC/{\it Spitzer} image at $3.6$ $\mu$m of the HII region RCW~34,
   outlining PAH emission,
   overlaid with contours of submm emission measured with LABOCA. Contours are
   0.06, 0.36, 1, 3, and 6 Jy/beam. 
                      }
              \label{rcw34:map}%
    \end{figure}

The location of the HII region \object{RCW~34} is enclosed within a box in the map of
submm emission from Vela C shown in Fig.~\ref{final:map}.
Figure~\ref{rcw34:map} displays an IRAC/{\it Spitzer} image of the region at $3.6$ $\mu$m
(from the GLIMPSE survey; Benjamin et al.\ \cite{benjamin}, 
Churchwell et al.\ \cite{churchwell}),
overlaid with contours of the submm emission.
The {\it Spitzer} image outlines
PAH emission excited by the ionising radiation from the massive stars. Clearly,
the submm emission is strictly related to the NIR 
diffuse emission, confirming that most of the submm structures in the 
field displayed in Fig.~\ref{rcw34:map} are associated with RCW~34.
There are indications that RCW~34 is farther away than Vela C; for example, the total radio flux
listed in the Parkes-MIT-NRAO catalogue at $4.85$ GHz (Griffith \& Wright \cite{parkes})
is $2.854$ Jy (source PMNJ0856--4305), about a factor ten fainter than 
that of the other H{\sc ii} region, RCW~36 (see below). The $V_{\rm LSR}$
of the main molecular clump is $\sim 5.5$ km s$^{-1}$, as measured in the CS(2--1)
transition (Bronfman et al.\ \cite{bronfy}). On the other hand, the $V_{\rm LSR}$
of the C$^{18}$O(1--0) clumps of Yamaguchi et al.\ \cite{yama} are $\sim 7$ km s$^{-1}$
in the northern and central parts of Vela C and only in the southern part are in
the range $5 - 6$ km s$^{-1}$.  Bik et al.\ \cite{bik10}
revised the distance of RCW 34 to $2.5$ kpc, which confirms that this is a background region.

\subsubsection{RCW~36}

\begin{figure}
   \centering
   \includegraphics[angle=-90,width=8cm]{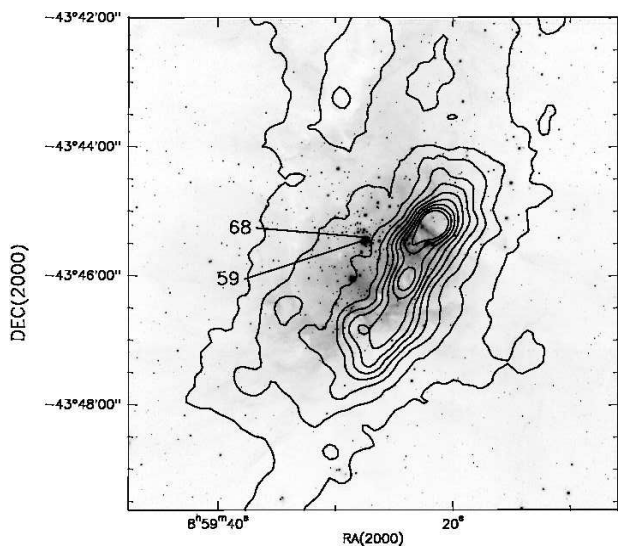}
   \caption{$K_{s}$ SofI image of the HII region RCW~36,
   overlaid with contours of submm emission mapped with LABOCA. Contours are
   in steps of $0.6$ Jy/beam from $0.2$ Jy/beam. The two late O stars
   that ionise the gas are labelled using the designation by Massi et
   al.\ \cite{massi03}.            
                      }
              \label{rcw36:map}%
    \end{figure}

The HII region \object{RCW~36} is one of the most prominent features of Vela C.
As shown in Fig.~\ref{final:map}, it is situated in the centre of
a large dust ridge and exhibits the most intense sources 
of submm emission in the field.
This region hosts a young star cluster and the IR source
IRS~34 studied by Massi et al.\ \cite{massi03}. The ionising sources are 
\# 68 and 59 of Massi et al.\ \cite{massi03}, which were identified by
those authors as two late O stars. Subsequently, Ellerbroek et al.\ \cite{elle13}
using optical-NIR spectroscopy confirmed that source \# 68
(their \# 1) is an O8.5--9.5 V star and \# 59 (their \# 3) is an O9.5--B0 V
star. 
As shown in Fig.~\ref{rcw36:map}, the submm emission is mostly 
arranged along a ridge, off-centred with respect to the cluster and
the ionising stars. Minier et al.\ \cite{minier} suggest that this actually marks an expanding ring of dust and gas surrounding the cluster that is tilted compared to the line of sight.
The Parkes-MIT-NRAO catalogue at $4.85$ GHz (Griffith \& Wright \cite{parkes})
lists a total flux density of 22 Jy from the H{\sc ii} region (source PMNJ0859--4345).

\subsubsection{IRS~31}

\begin{figure}
   \centering
   \includegraphics[angle=-90,width=8cm]{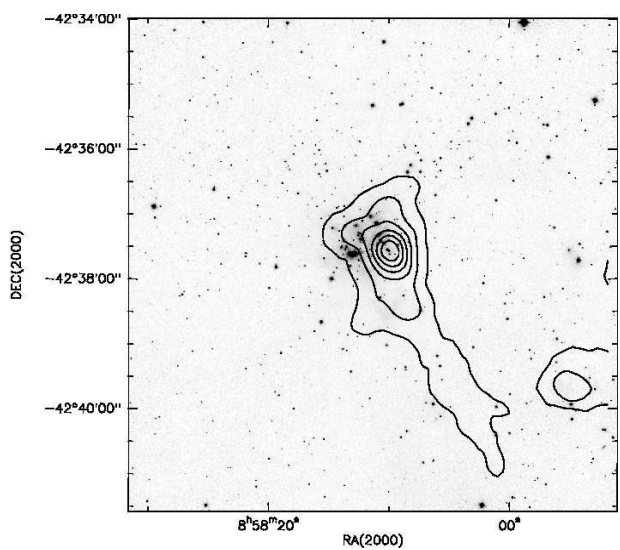}
   \caption{$K_{s}$ SofI image of the young embedded cluster IRS~31,
   overlaid with contours of submm emission mapped with LABOCA. Contours are
   in steps of $0.5$ Jy/beam from $0.5$ Jy/beam. 
                      }
              \label{irs31:map}%
    \end{figure}

\object{IRS~31} is a young embedded cluster discovered by Massi et al.\ \cite{massi03}.
It is located in the northern part of Vela C 
(see Fig.~\ref{final:map}) and, as can be seen in 
Fig.~\ref{irs31:map}, is associated with a strong submm source. Little is known about 
this cluster, whose most massive member is apparently a Class I source of
intermediate mass (Massi et al.\ \cite{massi03}). The NIR counterpart
of this object (\#14 of
Massi et al.\ \cite{massi03}) lies $\sim 30\arcsec$ east of the submm
peak. The dust emission peak appears off-centred compared to
the star cluster, as in the RCW~36 region. No radio emission at $4.85$ GHz
was detected by the Parkes-MIT-NRAO survey.   

%


\section{Observations and data reduction}
\label{obsec}
\subsection{LABOCA}
\label{labo}
The observations\footnote{Project ID: 087.F-0043 (MPIfR), 089.C-0744 (ESO)} 
were carried out in August 2011, May 2012, and August 2012, with the LABOCA camera 
(Siringo et al.\ \cite{Siringo2009}) at the APEX telescope
(G\"{u}sten et al.\ \cite{Gusten2006}). LABOCA is an array of 295 bolometers operating
at 345 GHz ($0.87$ mm) with a beam HPBW of $19\farcs2$.
The first set of observations (August 2011) is composed of small
($\sim 15\arcmin \times 15\arcmin$) on-the-fly maps covering the
Vela~C cloud, while the second set (May and August 2012) were obtained 
using single large maps in orthogonal coverages. 
Pointing and focus were checked regularly during the observations;
standard sources such as planets have been used as primary flux calibrators.

The data were reduced using the {\sl BoA} 
software\footnote{http://www.eso.org/sci/activities/apexsv/labocasv.html}
(Schuller \cite{schuller2012}),
following the steps described in detail by Schuller et al.\ \cite{schuller2009}.
The two datasets were then stitched together during the data reduction,
and emission from larger scales was iteratively recovered until convergence was reached
after four iterations. We checked that the source fluxes in the two datasets, when
reduced independently, are consistent with each other within a 30 \% level.

The final map was obtained using a pixel size of $4\farcs6$ after smoothing 
with a $9\arcsec$ FWHM Gaussian kernel. This covers the whole Vela C cloud as originally defined in
Murphy \& May \cite{MM} and subsequently imaged with {\it Herschel}
by Hill et al.\ \cite{hill11}. 
The r.m.s. typically ranges between 20 and 35 mJy/beam over the area, with the largest
values obviously found at the map edge. Figure~\ref{final:map} shows the
final dust emission distribution. 
Our observations clearly retrieved the structures 
already found through {\it Herschel} observations (Hill et al.\ \cite{hill11}). For
the sake of comparison, the 
sub-regions defined by Hill et al.\ \cite{hill11} are labelled in the figure,
along with the two HII regions RCW~34 and RCW~36, and the young embedded
cluster IRS~31.

\subsection{SIMBA}
Two regions in Vela C hosting young embedded clusters,
IRS~31 and RCW~36 (Massi et al.\ \cite{massi03}), 
were mapped in the 250 GHz ($1.2$ mm) continuum on May 26, 2003,
using the 37-channel SEST IMager Bolometer Array (SIMBA; Nyman et al. \cite{nyman}) at the Swedish-ESO
Submillimetre Telescope (SEST) located at La Silla, Chile. 
The observations were made at the end of ESO program 71.C--0088, 
whose objective was a large-scale map of Vela D
(Massi et al.\ \cite{massi07}). Two small maps of $600\arcsec \times
400\arcsec$ (azimuth $\times$ elevation) were obtained for both regions in fast scanning mode, 
with a scanning speed of $80\arcsec$ s$^{-1}$. The scan towards IRS~31 was repeated three times. 
The beam HPBW is $24\arcsec$ and the map r.m.s. is $\sim 30$ mJy/beam for RCW~36
and $\sim 20$ mJy/beam for IRS~31. Details on the data reduction and calibration steps are given in Massi 
et al.\ \cite{massi07}. These authors also found that their flux calibration is likely to be accurate
within $\sim 20$ \%.

\subsection{Clump-finding algorithms}
\label{retr:clump}

To obtain a census of the dense core population of Vela C, we used the algorithm CuTEx
(Molinari et al.\ \cite{mole}), developed to work on FIR/submm maps and widely tested on 
{\it Herschel} maps.
CuTEx computes the second derivative of the image, which is related to the curvature 
of the shape of the brightness distribution. 
The algorithm then searches for the strongest variations in brightness, 
which are expected to be at the centre of 
compact sources. Finally, all the pixels above a threshold limit are identified belonging to a 
candidate source. 
We tested different thresholds and selected the highest one which still retrieves all the 
features distinguishable by visual inspection. This choice minimises the number of artifacts. 
To strengthen our results we also discarded all the detections where the measured 
source peak { brightness} is 
smaller than three times the { brightness} r.m.s. around the source considered.
 
\begin{figure}
   \centering
   \includegraphics[angle=-90,width=8cm]{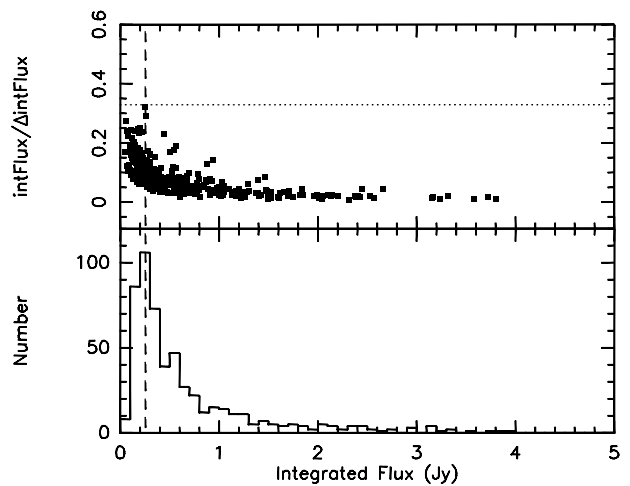}
   \caption{{\bf Top} Plot of the ratio of flux density
      to flux density error vs.\ flux density for all submm sources
      found with CuTEx; datapoints below the horizontal dotted line
      have $S/N > 3$; {\bf Bottom} Histogram of flux density for
      all submm sources found with CuTEx. The vertical dashed line marks
      the estimated completeness limit. 
                      }
              \label{err-plus-hist}%
    \end{figure}

We chose CuTEx on the grounds that it
is more focused on detecting compact sources
compared to other algorithms (e.g. CLUMPFIND) that generally split the (relevant) brightness distribution
into different, separated objects. Dust structures $\sol 0.1$ pc in size are usually
referred to in the literature as ``cores''. At a distance of 700 pc, this spatial scale corresponds
to angular sizes $\sol 30\arcsec$, resolved by LABOCA. Therefore,
hereafter we refer to the retrieved compact sources as submm cores.
We found 549 Vela C sources 
and a further 15 submm sources associated with RCW~34 (see Table~1).
As shown in Fig.~\ref{err-plus-hist}, most of the sources exhibit a signal-to-noise
ratio (S/N; in source {\em flux density}) much larger than 3. In particular, the histogram of 
flux densities, also shown in this figure, exhibits an increase in the number of sources
with decreasing flux density down to $F_{\rm compl} \sim 0.3$ Jy. This value corresponds to sources
with S/Ns of $\sim 5$ and can be interpreted as our completeness
limit for core detection, since the number of sources quickly decreases for flux
densities $< 0.3$ Jy. 
For this reason, we consider only sources above the completeness limit to derive 
the physical properties of the submm cores.

To allow comparison with other studies, we also tested the algorithm 
CLUMPFIND (Williams et al.\
\cite{willy}), which was run by Massi et al.\ \cite{massi07} on their SIMBA map (250 GHz) of
Vela D. We used
a lowest contour level of $\sim3$ r.m.s. and an intensity level increment of $\sim 5$ r.m.s.
as input parameters.
We performed several runs with different input parameters and found that this choice yielded the
best efficiency in terms of source detection based on a visual comparison of the results. 
We retrieved 291 Vela C submm sources and 7 other submm sources towards RCW~34 (see Table~2). 
Following
Massi et al.\ \cite{massi07}, we discarded another 104 CLUMPFIND detections with angular sizes
(not deconvolved)
smaller than the APEX beam size (see Table~2). However, similar detections in Vela D 
were classified as real faint sources by De Luca et al.\ \cite{deluke}
based on a cross-correlation of databases spanning a wide range of wavelengths. 

A visual comparison of the CuTEx and CLUMPFIND outputs shows that most of the
291 CLUMPFIND sources are associated with one or more CuTEx source. There
are only a few unmatched CLUMPFIND sources, but a significant number of unmatched CuTEx sources.
Interestingly, many of these unmatched CuTEx sources nevertheless are associated with
CLUMPFIND sources from the list of the 104 ``unresolved'' sources, confirming that the
latter are mostly real submm cores.  We therefore list
the 104 ``unresolved'' submm sources in Table~2 as well.
In the end, only a few CLUMPFIND and CuTEx sources do not match; in this respect the
two algorithms yield highly consistent results.

The estimated completeness limit $F_{\rm compl} \sim 0.3$ Jy translates into a {\it mass} 
completeness limit (Eq.~8 of Elia \& Pezzuto \cite{elia}, 
who use the formalism of Hildebrand \cite{hilde}) 
of $\sim 1.3$ $M_{\sun}$, by adopting a 
dust temperature $T_{\rm dust} = 15$ K, a dust opacity $k_{\nu} = 1.85$ g$^{-1}$ cm$^{2}$, and
a dust-to-gas ratio of 100 (Schuller et al.\ \cite{schuller2009}).
However, there are large uncertainties in the dust opacity, which may vary by more than 
a factor of two. In fact, if we assume $k_{\nu} = 0.5$ g$^{-1}$ cm$^{2}$ at 250 GHz, as 
in Massi et al. \cite{massi07} and in early seminal works like Motte et al.\ 
\cite{motte98}, and a dust emissivity index 
(see Table~\ref{name:def} for the
power-law index names and definitions adopted in this paper) 
 $\beta = 2$ (Olmi et al. \cite{holmes}), we determine 
$k_{\nu} = 0.86$ g$^{-1}$ cm$^{2}$ at 345 GHz (from Eq.~7 of Elia \& Pezzuto \cite{elia}).
Adopting this value and $T_{\rm dust} = 12.5$ K (in accordance with the discussion
in Sect.~\ref{cotemp}), 
the mass completeness limit increases to $\sim 3.7$ $M_{\sun}$.
As a sidenote, one should be aware that this is
a conservative choice: Martin et al.\ \cite{martin} reviewed the values of opacity
in the literature and the trend they found is clearly towards using a $k_{\nu}$ that is roughly 
a factor of two larger than that derived from the value used by Massi et al.\ \cite{massi07}.

%


\subsection{Ancillary data: WISE}
\label{wisedat}

Giannini et al.\ \cite{gianni12}  analysed 
{\it Herschel} observations of Vela C in the framework of the guaranteed time
key programme ``HOBYS''
(Motte et al.\ \cite{motte}).
Using CuTEx, these authors detected a number of compact sources, selecting a robust subsample of 256 
objects including both protostellar and starless cores. 
They identified starless cores out of their sample based on the lack of emission
at 70 $\mu$m and/or of associated IRAS/MSX/Akari IR sources. 
In this work, we alternatively exploit the association of CuTEx compact sources with 
IR sources from the WISE all-sky
data release (Wright et al.\ \cite{wise}). This provides point-source photometry at $3.6$,
$4.6$, 12, and 22 $\mu$m with better sensitivity and better angular resolution than IRAS, MSX, and
Akari. The latter ranges from $\sim 6 \arcsec$ at $3.6$ $\mu$m to $\sim 12 \arcsec$ at 22 $\mu$m.
We retrieved all point sources from the WISE catalogue that are projected towards the same sky
area as Vela C, with valid detections either in at least three bands or at 22 $\mu$m,
and with photometric errors $< 0.3$ mag either in all detected bands or
at least at 22 $\mu$m. Having valid detections in at least three
bands allows one to discriminate between field stars and real young stellar objects based on their
colours (see below). On the other hand, heavily embedded young sources might only be detected at 
22 $\mu$m due to their rising spectral energy distribution (SED).
First, we discarded all contaminants
(i.e. extragalactic objects, PAH emission, shock emission) by following the colour
criteria of Koenig et al.\ \cite{koenig}, from the sources with 
detections in every band needed to apply those criteria.
This means that all the sources lacking detection in any
of those bands are retained.  We then selected all remaining entries with 
the colours of Class I sources, again following  Koenig et al.\ \cite{koenig},
provided that all the needed colours are available from the WISE photometry. For sources
lacking detections in any of the bands required to apply the colour criteria, we computed 
their spectral index $\gamma$ 
(see Table~\ref{name:def} for definition)
and selected the ones with $\gamma > -0.3$
(Strafella et al.\ \cite{straf15}). Sources detected only at 22 $\mu$m 
with a photometric error $< 0.3$ mag were also retained.
Out of the 448 WISE Class I sources, only 22 were classified based on their
spectral index $\gamma$, and even less (four) are only detected at 22 $\mu$m with photometric 
error $< 0.3$ mag. 
Finally, we cross-correlated this subsample of WISE red sources and our CuTEx list of cores,
considering valid associations to be those where the WISE uncertainty ellipse overlaps the 
ellipse that has the major and minor FWHMs of a CuTEx output as its  semi-major and semi-minor axes.
Out of the 549 submm cores (excluding RCW~34),
217 were found to be associated with one or more WISE-selected Class I source
and were therefore classified as ``protostellar''. 
The remaining 332 submm cores were tentatively labelled as ``starless''.

One concern about WISE photometry is saturation of bright sources in regions of intense diffuse
emission. A look at the archived images shows that only a small area towards RCW~36
is saturated in the two upper bands. This area includes 12 submm sources.
We checked the MSX Infrared Point Source Catalogue
for IR sources with $\gamma > -0.3$ matching any of the 12 sources in projection.
Thus, we found another 
four cores with associated red MSX sources, which were reclassified as protostellar. 

    \begin{figure*}
    \centering
    \includegraphics[angle=-90,width=18cm]{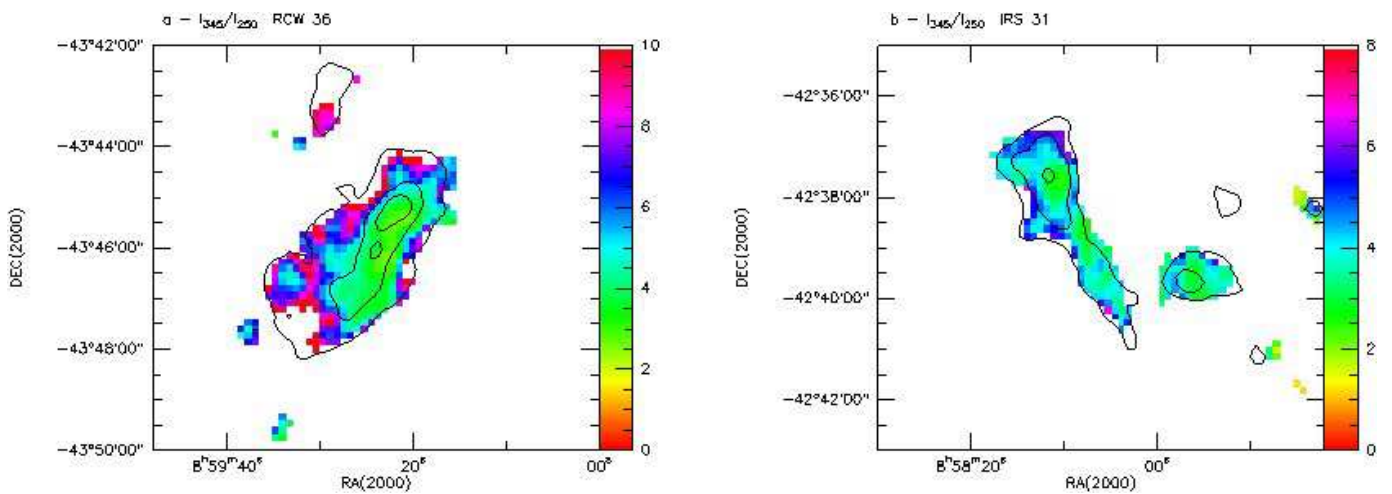}
    \caption{Colour map of $S_{345}/S_{250}$ towards {\bf a} RCW~36 and {\bf b} IRS~31.
       The map at 345 GHz (LABOCA) has been smoothed to the angular resolution of
       the map at 250 GHz (SIMBA) and then divided by it. 
          The emission at 345 GHz is overlaid in contours of {\bf a} 1000, 3000, and
       5000 Jy/beam and {\bf b} 500, 1000, and 3000 Jy/beam. Only pixels with
       a brightness $\protect \sog 5$~r.m.s. have been selected. 
                       }
               \label{flux:ratio}
     \end{figure*}
%

\subsection{Ancillary data: Herschel}
\label{anc:fir}

Dust emission at submm wavelengths is optically thin even for large column densities
(e.g. Schuller et al.\ \cite{schuller2009}),
allowing a straightforward determination of gas mass
under the assumption of a gas-to-dust mass ratio.
Dust temperature is also needed. The availability of 
{\it Herschel} observations between 70 and 500
$\mu$m, with a beam size ($18\farcs1$ at 250 $\mu$m) comparable with the APEX one
at $345$ GHz, provides us with a means of deriving reliable dust temperatures. We 
decided to cross-correlate the sample of {\it Herschel} cores from Giannini et al.\
\cite{gianni12} and our CuTEx-extracted sample to infer dust temperatures for 
the submm cores.   

We retrieved the 
{\it Herschel}
maps at 70/160/250/350/500 $\mu$m observed in the HOBYS
key programme and used CuTEx with the same parameters as in Giannini et al.\ 
\cite{gianni12}.  The compact sources found in each {\it Herschel}
map were matched following Giannini et al.\ \cite{gianni12} to re-build their catalogue
of 1686 entries, most of them with a detection in one or two bands only.
A set of routines was provided by Giannini (private communication) to
make sure that we could quickly reproduce their results. 
These sources were then matched to the submm cores retrieved with CuTEx
by requiring that
a circle of the same diameter as the beam at 250 $\mu$m and centred at the
nominal position of the 
{\it Herschel} source overlaps at least partially with the CuTEx
ellipse (with axes of 1 FWHM) of a submm core. In quite a few cases, more 
{\it Herschel} sources
were associated with the same submm core; we selected the closest match, discarding 
the others. 
The reverse case (multiple submm cores associated with one 
{\it Herschel} source)
occurred on a few occasions, and we selected the closest match as well. 
Ultimately, we found that 219 out of 328 submm starless cores 
and 185 out of 221 submm protostellar cores are associated with 
{\it Herschel} compact sources,
meaning that we were only able to match 74 \% of the submm cores to 
{\it Herschel} sources. By overlapping CuTEx ellipses, and IR and submm maps,
we found that submm sources without a {\it Herschel} counterpart are either part
of a multiple association with the same {\it Herschel} source, or fall towards
250 and 350 $\mu$m emission which is probably too extended and hence missed by 
CuTEx in those maps. 
A small number of other submm sources without a 
{\it Herschel} match
may be artifacts, but we checked that in this case their flux density is fainter 
than our completeness limit.

Only 388 {\it Herschel}
sources passing the first three selection criteria adopted by Giannini 
et al.\ \cite{gianni12} were suitable to derive a temperature. Therefore, in the end,
the subsample of those {\it Herschel} compact sources with a temperature
determination and matched to submm cores includes only 197 objects (i.e. 51 \%
out of 388 {\it Herschel} entries).
By overlaying the positions of {\it Herschel} and submm sources,
we found that only 3 out of the 48 protostellar sources of Giannini et al.\ \cite{gianni12} 
do not have a matching submm source. It is therefore the starless 
{\it Herschel} sample of these latter authors that exhibits the
higher fraction of unmatched sources. 
A comparison between the {\it Herschel} catalogue and our extraction list reveals that we 
detected more sources over the filamentary features present in the {\it Herschel} maps. 
This is likely due to the approach adopted when treating the {\it Herschel} maps, 
that is, using a higher threshold value over the bright filamentary features of the clouds 
to prevent a fragmentation of the emission in an implausibly large number of individual sources.

The lists of identified sources from the {\it Herschel} and LABOCA maps are therefore 
partly different. CuTEx also tends to find
larger cores on the submm map, whereas these are often split up into a number of smaller
sources on the IR maps. This is due to both 
the differences (in sensitivity) between the two datasets and the slightly different 
approaches adopted in defining the parameters for the extraction.

\section{Results}
\label{resec}

The derived statistical properties of the population of submm cores in Vela C
are described in the following. The involved physical parameters
are also discussed along with the main sources of error.

\subsection{Nature of the submm emission near young star 
clusters and H{\sc ii} regions}

The nature of the submm sources in the most evolved regions of Vela C,
namely the young embedded clusters and H{\sc ii} regions, can be further investigated
by comparing our LABOCA data at 345 GHz and SIMBA observations of IRS~31 and
RCW~36 at 250 GHz. Figure~\ref{flux:ratio} shows the pixel-to-pixel ratio
of flux density at 345 and 250 GHz after smoothing of the LABOCA map to the 
spatial resolution of the SIMBA map ($24 \arcsec$). This ratio 
gives some useful information on the dust spectral index, provided contamination
from optically thin free-free emission (which would yield a ratio $\sim 1$) is
negligible.

Both maps of Fig.~\ref{flux:ratio}
show that the flux density ratio 
is $\sim 2.7$ towards the centre of each emitting clump 
and increases towards the outer parts (up to $5-6$ for RCW~36 and $4-5$
for IRS~31). This gives some hints on the spatial variations of
dust emissivity index and temperature. In principle, an increase
in the ratio towards the outer parts may be a signature of a temperature gradient,
with the innermost clump regions being colder. This is plausible, particularly 
for RCW~36 where the clump is externally exposed to an intense UV field from massive stars.

A ratio of $\sim 2.7$ can be obtained (from Eq.~6 of
Elia \& Pezzuto \cite{elia}) with a dust emissivity index
$\beta$
ranging between $1.3$ and $1.9$ for $T_{\rm dust}$ between 10 and 30 K, whereas a
ratio of $\sim 4$ requires either $\beta \sim 2.5 - 3.2$ in the same temperature
range or a dust temperature exceeding 100 K with $\beta \sim 1-2$. 
Larger values of the flux density ratio ($> 5$) require $\beta > 3$.
Thus, 
the innermost emission from both clumps is consistent with $\beta \sim 1-2$, the
range of values expected in dust cores (see, e.g. Schnee et al.\ \cite{schnee}). 
This agrees with extrapolating the total Parkes-MIT-NRAO radio flux density at $4.85$ GHz from
RCW~36 to 345 GHz in the assumption of optically thin emission, which yields
a value that is only $\sim 10$ \% of the total 345 GHz flux density from a similar area.
On the other hand, an increase in the dust temperature of the outermost clump layers can
only explain part of the increase in flux density ratio, as this would also need
unrealistic $\beta$ values. Thus, systematic effects may be involved. 
Position errors (few arcsec) would mostly affect the ratio in the clump edge
where the emission is fainter. Alternatively, it is possible that the flux
density at 250 GHz has been systematically underestimated in the less intense emission areas. While reducing these data
(following Massi et al. \cite{massi07}), we noted some small changes in 
the source morphology depending on the choice of the source masking area.
In turn, this suggests that low-level mm emission has not been fully
recovered in our 250 GHz map. A similar effect was noted by Massi et al. \cite{massi07}
when comparing photometry of sources from their large-scale map of Vela D and
from much smaller maps provided by other authors.

%
%
\addtocounter{table}{2}
\begin{table}
         \caption{Names and definitions used in the text for power-law indices.
\label{name:def}}      
\scriptsize
\centering                          
\begin{tabular}{l | l | l }        
Name & Definition & Astrophysical quantity \\
\hline\\
$\alpha$ & $\alpha = {\rm d}\log(\psi)/{\rm d}\log(M)$ & $\psi =  {\rm d}N/{\rm d}M$ 
initial mass function \\
$\alpha'$ & $\alpha' = {\rm d}\log(\psi')/{\rm d}\log(M)$ & $\psi' =  {\rm d}N/{\rm d}\log(M)$ 
initial mass function \\
$\beta$ & $\tau_{\nu} = (\nu/\nu_{0})^{\beta}$ & $\tau_{\nu}$ dust column density \\
$\gamma$ & $\gamma = {\rm d}\log(\lambda F_\lambda)/{\rm d}\log(\lambda)$ &
$\lambda F_\lambda$ spectral energy distribution \\
\hline
\end{tabular}
\end{table}

We can conclude that even in the
H{\sc ii} region RCW~36, where regions with intense free-free emission are expected,
the flux density ratio clearly indicates that the submm emission detected
with LABOCA is dominated by dust
thermal radiation with $\beta \sim 1-2$. External heating is also plausible,
although our results are likely to be biased by systematic errors.

\subsection{Statistical properties of the submm cores}
\label{statcores}

In this section we discuss a few general properties of the submm cores by analysing
the results obtained with both CuTEx and CLUMPFIND. 
The gas masses were derived for the starless and protostellar subsamples 
from Eq.~8 of Elia \& Pezzuto \cite{elia} with
a dust opacity $k_{\nu}(\nu = 345 {\rm GHz}) = 0.86$ g$^{-1}$ cm$^{2}$
(see Sect.~\ref{retr:clump}). We adopted this value for immediate comparison
of the derived masses with those obtained by Massi et al.\ \cite{massi07}
in Vela D. To be consistent with the ATLASGAL catalogue 
(Schuller et al.\ \cite{schuller2009}), 
our masses have to be scaled by a factor of
$0.46$. The single-temperature approximation
of Sect.~\ref{cotemp} is used for an easy comparison.

\begin{figure}
   \centering
   \includegraphics[angle=-90,width=8cm]{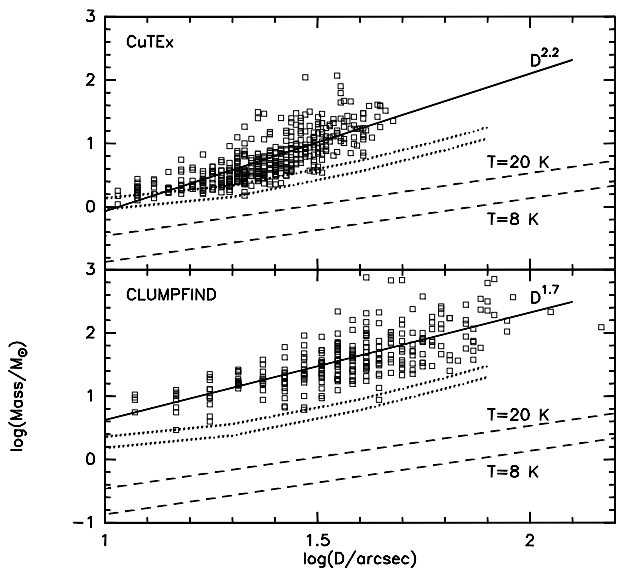}
   \caption{Mass vs.\ size for the submm sources in Vela C. The mass is
     computed by assuming a constant $T_{\rm dust} = 12.5$ K. The cores
     retrieved with CuTEx are plotted in the top box, whereas the ones found
     with CLUMPFIND are plotted in the bottom box. The solid line is
     a linear fit (in the log-log space) to the datapoints.
     The dotted lines are the detection limits estimated assuming r.m.s. values 
     of 20 and 30 mJy/beam (see text).
     Finally, the dashed lines are the theoretical loci of Bonnor-Ebert critical
     half-masses at $T = 8$ K and $T = 20$ K.
                      }
              \label{mass:size:fig}%
    \end{figure}

Following Massi et al.\ \cite{massi07}, we examined the relationship between mass and
size in the two samples, which is shown in Fig.~\ref{mass:size:fig}. The maximum
and minimum source sizes have been deconvolved and their geometrical average is displayed
({\it x} axis). Only sources with deconvolved apparent mean diameter larger than half the
beam size have been plotted. From a linear fit
in the log-log space, we found $M \sim D^{1.7 \pm 0.1}$ for
CLUMPFIND sources and a
slightly steeper relation ($M \sim D^{2.2 \pm 0.1}$) for CuTEx sources. The various biases
affecting the results of different clump finding algorithms have long been known
(e.g. Schneider \& Brooks \cite{kate}, Curtis \& Richer \cite{cur:ric}); in this
respect, CLUMPFIND and CuTEx yield similar mass--size relations. In addition, the CuTEx datapoints
spread over the same plot region as that of the 
{\it Herschel} data of Giannini et al.\
(\cite{gianni12}; see their Fig.5), 
although the latter do not exhibit a clear trend of increasing mass with
size. As the detection limit depends on the apparent source size, 
detection loci can be derived as follows. 
By taking into account typical values of 
r.m.s. of 20 and 30 mJy/beam (see Sect.~\ref{labo}) and the pre-selection limit of
three times the r.m.s. used for the CuTEx output, in the same assumptions as above, $0.06$ 
Jy and $0.09$ Jy
correspond to detection
limits of $0.7$ M$_{\sun}$ and $1.1$ M$_{\sun}$, respectively, for
{ point sources} with $T_{\rm dust} = 12.5$ K. 
For extended sources with a 2D Gaussian
spatial distribution of size $D$, this limit has to be scaled by $(B^{2} + D^{2})/B^{2}$,
where $B$ is the beam size.
As for CLUMPFIND, the detection loci have been computed assuming a detection limit
of 5 r.m.s., more consistent with the parameters used.
These loci are highlighted in Fig.~\ref{mass:size:fig} by dotted lines. The more massive
cores are clearly the bigger ones, although the lower envelope is affected by
incompleteness. Therefore, the derived slopes may be in error and the mass--size
relations might be slightly steeper. 

%


Following Giannini et al.\ \cite{gianni12},
Figure~\ref{mass:size:fig} also displays the theoretical
loci of Bonnor-Ebert half critical masses ($M_{\rm BE}$)
for $T = 8$ K and $T = 20$ K. Most of the cores clearly lie in
the plot region where gravity dominates over thermal support. Although 
turbulence and magnetic field will provide further support,
they are only likely to delay the collapse of cores that have not yet
formed a protostar. This does not necessarily apply to cores hosting protostars,
in which increased turbulence from outflows may eventually disrupt their structure. 

\begin{figure}
   \centering
   \includegraphics[angle=-90,width=8cm]{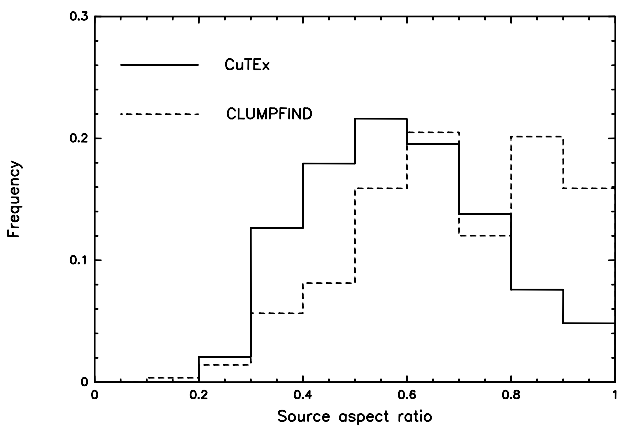}
   \caption{Normalised distribution of ratios of minimum to maximum source size
     for the cores found by CuTEx (solid line) and the ones found by
     CLUMPFIND (dashed line). Only sources with minimum deconvolved size 
     larger than half the LABOCA beam have been taken into account. 
                      }
              \label{hist:size:ratio}%
    \end{figure}

In view of the filamentary structure of the cloud, it is of interest to have a look at 
the distribution of ratios of minimum-to-maximum deconvolved source size, shown in 
Fig.~\ref{hist:size:ratio}. The peak between $0.5$ and $0.6$ suggests that the smallest
structures tend to be prolate rather than filamentary.
As shown in Fig.~\ref{hist:size:ratio}, the distribution of deconvolved source size ratios 
of the CLUMPFIND cores appears even less
filamentary than that of the CuTEx cores. By plotting the distribution of the position angles
of the CuTEx sources in the south, centre, and north parts of the cloud, we were unable to find
any significant preferred orientation of the submm cores, which would also have pointed to a
filamentary nature. 

The total mass in dense cores retrieved by CuTEx is
$\sim 5700$ $M_{\sun}$, whereas the total core mass retrieved by CLUMPFIND is $\sim 10000$ $M_{\sun}$. This
is not unexpected because unlike CLUMPFIND, CuTEx is optimised for searching compact sources.
Further, CLUMPFIND is able to include more low-level emission in a clump
(see Fig.~\ref{mass:size:fig}; also Cheng et al.\ \cite{cheng}). Taking into account
a total mass of $5.3 \times 10^{4}$ $M_{\sun}$ from $^{13}$CO emitting gas for Vela C 
(Yamaguchi et al.\ \cite{yama}),
this indicates that 10--20 \% of the gas is in dense cores. This figure is consistent with 
the low star-formation efficiency of GMCs (Padoan et al.\ \cite{padoan})
and the typical dense gas fraction in the inner Galaxy (e.g. Csengeri
et al.\ \cite{csen}). 

%

\begin{figure}
   \centering
   \includegraphics[angle=-90,width=8cm]{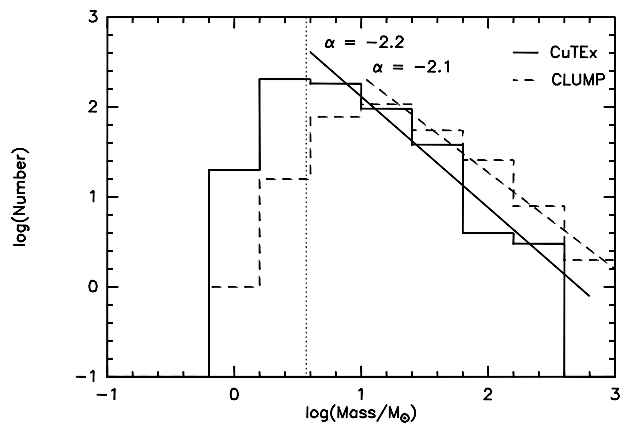}
   \caption{Core mass function for the CuTEx sources (solid line) and
     the CLUMPFIND sources (dashed line). Also shown are linear fits
     to the CMFs corresponding to a relation ${\rm d}N/{\rm d}M \sim M^{-2.2}$
     (CuTEx CMF; solid line) and ${\rm d}N/{\rm d}M \sim M^{-2.1}$
     (CLUMPFIND CMF; dashed line), respectively. The vertical dotted line
     marks our estimated completeness limit.
                      }
              \label{cTdust:mass:hist}%
    \end{figure}

The mass distributions, or CMFs, 
are plotted in Fig.~\ref{cTdust:mass:hist} for the CuTEx and CLUMPFIND samples.
For both of them
the CMF increases with decreasing mass down to a peak roughly located at the
completeness limit of $M \sim 3.7$ $M_{\sun}$ (dotted line in figure),
as estimated in Sect.~\ref{retr:clump}.  A linear $\chi^{2}$ fit (in the
the log-log space) to this part of the CMF
yields a slope of $\alpha' = -1.2 \pm 0.2$ (CuTEx) and $\alpha' = -1.1 \pm 0.2$ (CLUMPFIND),
where $\alpha' = \alpha - 1$ and $\alpha = 2.35$ for a Salpeter IMF 
(Salpeter \cite{salpeter}; see Table~\ref{name:def} for 
definitions).
We note that the histogram peaks have not been included in the fitted data range to
minimise incompleteness effects.

As can also be deduced from Fig.~\ref{mass:size:fig}, it appears that
CuTEx tends to further subdivide the largest CLUMPFIND sources in smaller cores.
Nevertheless, the power-law indices $\alpha$ are roughly consistent with each other
at a $1 \sigma$ level.
In the following we only discuss the results from the CuTEx algorithm
without explicit mention. 
In addition, the submm cores 
located inside the sky area shown in Fig.~\ref{rcw34:map} (hence likely to be
associated with RCW~34) will not be taken into account.

%


\subsection{Starless and protostellar cores}
\label{s:p:c}

A further step in our analysis consists in separating the CuTEx sources into starless and protostellar
cores and comparing the properties of the two subsamples. In principle, after 
gravitationally bound starless
cores have evolved into protostellar cores they will lose part of their mass due to 
accretion onto the central object(s) and outflowing activity, developing temperature 
gradients as well. The two populations
are therefore expected to exhibit different physical properties. In addition, if
star formation proceeds in bursts, less evolved regions should display a larger fraction
of starless cores, in contrast to more evolved regions. On the other hand, if 
star formation is continuous, the number of starless and protostellar cores will be related to the
lifetimes in each stage.

Based on their association with WISE (and MSX) point sources, as explained in Sect.~\ref{wisedat},
we classified the submm sources into starless cores (328) and 
protostellar cores (221); a further refinement is explained in the following section.  
We note that cross-matching of LABOCA and WISE yielded a larger fraction of protostellar cores
than Giannini et al. \cite{gianni12} found in their {\it Herschel} sample, that is, 
40 \% of 549 submm cores versus\ 18 \% of 268 
{\it Herschel} cores.

Deriving the completeness limits of the WISE photometry is mandatory to assess the reliability of our
catalogue of starless cores. We examined the histograms of the number of  mid-infrared (MIR) sources versus  magnitude;
taking into account the effects of the cuts required to fulfil the criteria of Koenig et al.\ 
\cite{koenig},
rough completeness limits are $[3.6] \sim 14$, $[4.6] \sim 12$, $[12] \sim 9$ and $[22] \sim 7$.
These values are 1--3 mag brighter than the sensitivity limits quoted in 
the WISE Explanatory Supplement\footnote{http://wise2.ipac.caltech.edu/docs/release/allsky/expsup/}
for the relevant sky region.
Once converted into flux units and, for example, compared with the models of Class I and Class 0 sources
of $0.5 M_{\sun}$ by Whitney et al.\ \cite{whitney}, it can be seen that the completeness limits at 
$3.6$ and $4.6$ $\mu$m are faint enough to detect such objects taking into account a
distance of 700 pc and a further foreground reddening up to $A_{V} =20$. Even in the worst case of
edge-on discs, these objects would be detectable at $3.6$ and $22$ $\mu$m. 
Furthermore, the completeness
limit at 22 $\mu$m is faint enough to allow detection of Class I and Class 0 sources of even-lower-mass central objects. 
Alternatively, one can compute the bolometric luminosity following Kryukova et al.\
\cite{kryukova}. Starting from our completeness limit at 22 $\mu$m, after
conservatively dereddening it by $A_{V} = 20$, we assumed a spectral index
$\gamma = -2$ 
(see Table~\ref{name:def} for definition)
to compute the MIR luminosity from Eq.~6
of Kryukova et al.\ \cite{kryukova}. Equation~7 of Kryukova et al.\ \cite{kryukova} then 
yields $L_{\rm bol} \sim 1.7 - 2.8$ $L_{\sun}$,
depending on whether the NIR flux is neglected (which may be the case) or
extrapolated from $\gamma = -2$.  A comparison
with the birthline of Palla \& Stahler \cite{pa:stah} indicates a mass
of $\sim 0.4-0.5$ $M_{\sun}$ for the central protostar.
For the sake of comparison, we can roughly estimate the completeness limit in
central masses of the
{\it Herschel} protostellar cores in Giannini et al.\ \cite{gianni12} using their quoted
completeness limit at 70 $\mu$m of $0.21$ Jy and following Dunham et al.\ \cite{dunham}.
By using Eq.~2 of Dunham et al.\ \cite{dunham}, scaled to a distance of 700 pc, 
we found that the flux density at 70 $\mu$m
translates into a bolometric luminosity of the central (proto)star $L_{\rm bol} \sim 0.28$ 
$L_{\sun}$ (we note that Dunham et al.\ \cite{dunham} indicate this luminosity as $L_{\rm int}$). 
 We highlight the fact that the 70 $\mu$m emission is in principle a more sensitive  
protostellar tracer than WISE. However, this contrasts with the much lower number of protostellar
cores found by Giannini et al. \cite{gianni12}, which may be due to a poorer effective 
sensitivity because of their selection criteria. 

It is important to note that in principle the sample of starless cores will contain both
transient unbound gas structures and real prestellar gas condensations. By considering the
limited range of dust temperatures envisaged for the starless cores (see Sect.~\ref{cotemp}),
a comparison between the location of the submm datapoints and the theoretical loci
of Bonnor-Ebert critical half-masses $T = 8$ and $20$ K  in Fig.~\ref{mass:size:fig}
suggests that most of the submm
starless cores are likely to be gravitationally bound. Therefore, although we refer to them
as ``starless'', they are actually likely to be prestellar in nature. 
 
\subsection{Core temperatures}
\label{cotemp}

Given that the submm sources are in 
different evolutionary stages, and therefore have different temperatures, 
one should use the correct temperature for each object for an accurate mass determination. 
Based on Giannini et al.\ \cite{gianni12}, a few general trends are well established. 
These latter authors found a very narrow interval of dust temperatures,
with means of $\sim 10$ K and  $\sim 13$ K for their starless and protostellar cores, respectively. In particular, the temperature distribution for starless
cores is strongly peaked at $\sim 10$ K, with a cut-off at $\sim 15$ K. Conversely,
the temperature distribution of protostellar cores appears almost uniform between 9 and 
15 K, with a small number of warmer sources (up to $\sim 25$ K). Netterfield et al.\
\cite{Nett09} analysed a much larger sample of sources using BLAST data
(250, 350, and 500 $\mu$m) from the whole ($\sim 50$ deg$^{2}$) Vela region. They 
found a temperature distribution ranging between 10 and 15 K for cores without
associated IRAS or MSX sources, similar to that of the starless sample of Giannini et al.\ 
\cite{gianni12}. However, the temperature distribution for all BLAST cores
peaks at $\sim 12.5$ K with 19 \% of the sources warmer than 20 K.
This is also confirmed by the temperature distribution
derived by Hill et al.\ \cite{hill11} from the 
{\it Herschel} observations of Vela C. These latter authors found
a bimodal temperature distribution, with a peak at $\sim 15$ K and a tail extending
to $\sim 30$ K. Remarkably, the southernmost regions are characterised by a strongly
peaked distribution (with peak temperatures $< 15$ K), and only the Centre-Ridge
and the northernmost part exhibit a small fraction of warmer regions. 
Fissel et al.\ \cite{fissel} found (from the BLAST data) a number of lines of sight
where dust appears heated by the HII region RCW~36 in the central part of Vela~C. 

We take two different approaches to assign a dust temperature to each core.
A first-order approximation widely used in the literature consists in assuming a single
temperature for all sources. Given the
small range of temperatures involved as discussed above, 
particularly for starless cores, a single value
should represent a reasonable assumption.  We chose $T_{\rm dust} = 12.5$ K for this
first-order approximation. This value would underestimate the
mass of cores with $T_{\rm dust} = 10$ K by only 35 \%, and overestimate the mass of cores with $T_{\rm dust} = 15$
K by only 37 \%. 
Undoubtedly, a number of warm sources ($T_{\rm dust} > 15 - 20$ K) 
are expected in our sample, but many of them are probably located in areas  
near stellar sources radiating intense UV fields, such as the HII region
RCW~36, and represent only a small percentage of the total. 

The second approach consists in deriving single-core dust temperatures for the
197 {\it Herschel} compact sources associated with submm cores as described 
in Sect.~\ref{anc:fir}. For each {\it Herschel} source
we performed flux scaling at the various 
{\it Herschel} wavelengths (a method adopted by Motte et al.\ \cite{motte})
and then fitted a greybody to the SED, both explained in 
Giannini et al.\ \cite{gianni12}. The submm points are not included in the fit. 
We also note that the {\it Herschel} HPBW is
larger than the one of LABOCA for $\lambda \geq 350$ $\mu$m ($\sim 25 \arcsec$
at $350$ $\mu$m and $\sim 36 \arcsec$ at 500 $\mu$m). In principle, unresolved
cold dust structures might bias the derived temperature, but this should be mitigated
by the flux scaling in most cases.  
At the end of the process, we were
able to derive a dust temperature for 86 of the
328 submm starless cores and 111 of the 221 submm protostellar cores of our sample.

\begin{figure}
   \centering
   \includegraphics[angle=-90,width=8cm]{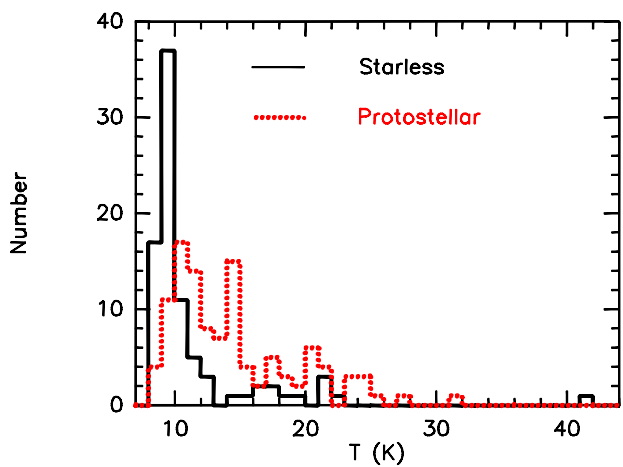}
   \caption{Histograms of the dust temperatures of starless (solid black line)
     and protostellar (dotted red line) submm cores as derived from their FIR 
     {\it Herschel} counterparts.
                      }
              \label{histo:temp}%
    \end{figure}

Figure~\ref{histo:temp}
shows single-core temperature histograms for the submm starless and protostellar cores. The two
distributions are clearly different, with starless cores exhibiting a strong peak
at $\sim 9 $ K and protostellar ones peaking at $\sim 11$ K (the other peak at $\sim 15$ K
disappears after shifting the bin centres by $0.5$ K)
and exhibiting a significant tail at higher temperatures.
A look at the 12 starless cores with a temperature $> 14$ K indicates that all
but one have been detected at 70 $\mu$m and suggests that these are probably 
protostellar sources contaminating the starless sample. This further constrains
the temperature of the starless cores in the range 9--12 K and enhances
the difference between the two distributions, confirming that the two samples
are different in nature. A comparison with Fig.~4 of Giannini et 
al.\ \cite{gianni12} shows that our temperature distributions are
not very different from theirs, although
their statistics are probably biased by their much smaller sample of protostellar objects.

We note that the limited number of submm sources associated with a 
{\it Herschel} source 
prevents us from using the whole sample of submm cores.
However, if the temperature distribution of starless cores is representative of
all starless cores in the region, using the right temperature for each source
is not strictly necessary. In this respect, the single-temperature approach allows
us to use the whole submm sample to explore its statistical properties.

\begin{figure}
   \centering
   \includegraphics[angle=-90,width=8cm]{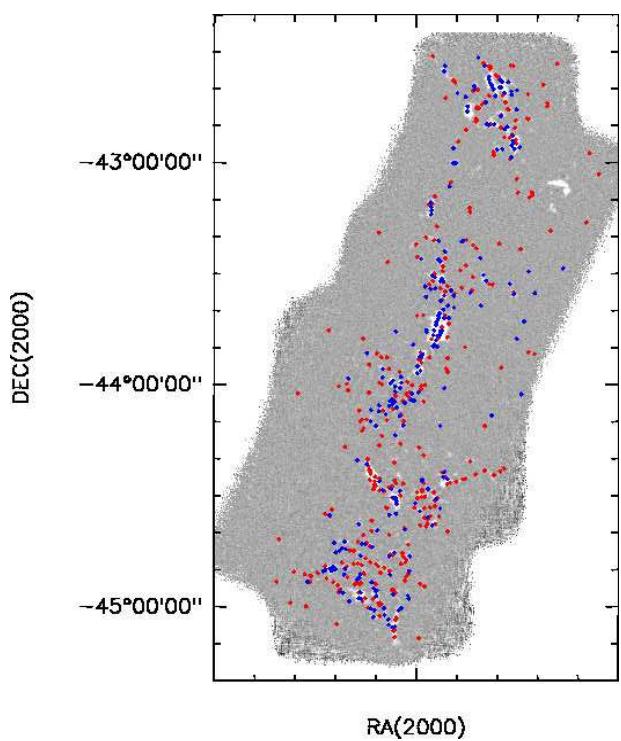}
   \caption{Projected spatial distribution of starless cores (red diamonds)
      and protostellar cores (blue diamonds) overlaid on the
        LABOCA map of 345 GHz emission from Vela C.      
                      }
              \label{dist:pre:proto}%
    \end{figure}

In the end, the 12 starless sources with
$T > 14$ K were re-classified as protostellar. Therefore, the final subsamples are
composed of 316 starless and 233 protostellar cores. 
The spatial distribution of starless and protostellar cores is shown in Fig.~\ref{dist:pre:proto}.
We use the same symbols as Giannini et al. \cite{gianni12} for comparison with their
Fig.~1. 
It is also interesting to derive the percentage of protostellar 
cores in different sky areas. This confirms the visual impression that the southern
part of Vela C hosts the lowest fraction of protostellar cores ($\sim 35$ \% at ${\rm DEC} < 
-44\degr 15\arcmin$),
while the central part hosts the largest fraction ($52$ \% at $-44\degr 15\arcmin
< {\rm DEC} < -43\degr 10\arcmin$).
At ${\rm DEC} > -43\degr 10\arcmin$, $\sim 41$ \% of the submm cores are protostellar. 
 
%


\subsection{Core masses: individual and statistical properties}
\label{c:m:f}

The masses of the submm cores were derived from their 345 GHz flux density by
using Eq.~8 of Elia \& Pezzuto \cite{elia}, with a dust
opacity $k_{\nu} = 0.86$ g$^{-1}$ cm$^{2}$
(see Sect.~\ref{retr:clump}). Dust temperatures were assigned based on
the two approaches outlined in Sect.~\ref{cotemp}, and we
assumed a gas-to-dust ratio of 100. 

Errors on the mass estimates are due to uncertainties on distance, dust opacity, temperature, and flux
calibration. The effects of distance errors are tabulated in the online Appendix. 
Dust opacity is discussed by Ossenkopf \&
Henning \cite{osse} who conclude that uncertainties are 
unlikely to exceed a factor of approximately two, even when different physical environments
are taken into account. 
Having selected the same molecular cloud and specific classes of objects
(namely prestellar and protostellar cores), this should not be the case
within each of the classes. These authors show that,
providing cores remain stable over a timescale of $\sim 10^{5}$ yr, opacity gains only 
a little dependence on gas density due to dust coagulation.
The mass--size relation
that we find for all cores implies that their mean gas densities range within an order of magnitude,
that is, $\sim 10^{5}$ -- $10^{6}$ cm$^{-3}$. It can be
seen from Ossenkopf \& Henning (\cite{osse}; 
their Tables 2 and 3) that opacity variations are less than
30 \% for densities in the range  $10^{5} - 10^{6}$ cm$^{-3}$
(less than 20 \% for densities of $10^{6} - 10^{8}$ cm$^{-3}$) and dust grains with ice mantles.
One might suppose that the main error contributed by dust opacity would result in a zero-order term which would translate into a mass scale factor
(probably $\sol 2$), as for distance errors. This can be caused by an incorrect choice of
the most suitable dust models for the class of objects considered.
We also note that the scaling factors to use if simultaneously adopting
the ATLASGAL dust opacity and the revised distance of 950 pc would partly cancel each other. 
Flux calibration errors
are more likely to affect all sources in a relatively large cloud area
in the same way, rather than yield random source-to-source flux fluctuations. 
As stated in Sect.~\ref{labo}, we found that the fluxes of the submm sources are consistent within 30 \%
when observed in different runs. Having mapped most of Vela C a few times, the actual 
calibration accuracy should be better than that. However, even a mass
variation of 30 \% due to calibration or dust emissivity errors turns into a
variation of the logarithm of mass of $\sim 0.1$, which has to be compared, for example, with the 
logarithmic size of the bins we used in the mass histograms: $0.4$.
In conclusion, although core masses may even be  systematically incorrect by up to a factor  of approximately two, 
parameters such as 
the CMF slope are unaffected by distance uncertainties and are only very mildly affected by
dust opacity or calibration uncertainties; this is especially true for prestellar cores which
exhibit only small temperature differences and are not expected to have inner
temperature gradients (making their physical environment relatively constant).
The derived masses also depend on the adopted dust temperature and, as seen above, errors
of a few Kelvin cause significant changes. 
The effects of temperature on the CMF are therefore discussed in more detail later. 

Table~\ref{core:prop} compares the statistical properties of the various samples of sources,
namely  the {\it Herschel} FIR sources from Giannini et al.\ \cite{gianni12}, 
the LABOCA submm sources with temperature derived from matching them to 
{\it Herschel} sources, and the LABOCA submm sources with an assumed constant temperature.
Each submm subsample (starless and protostellar) clearly exhibits similar properties irrespective of the
assumed temperature. The most notable difference
lies in the high-mass end of the mass distributions. We checked that only 7 submm protostellar
sources have a mass $> 50$ $M_\sun$ when computed assuming $T_{\rm dust} = 12.5$ K. Six of them
are included in the sample of submm protostellar sources with temperature from 
{\it Herschel} data, but having been
assigned a higher $T_{\rm dust}$ they exhibit lower masses here. 
On the other hand, only two submm starless sources
have a mass $> 40$ $M_\sun$ when computed assuming $T_{\rm dust} = 12.5$ K. Neither of them is
included in the sample of submm starless sources with a temperature from 
{\it Herschel} data. Both are
located in the area of RCW~36, a region with a high surface density of submm sources.

%
%
\begin{table*}
         \caption{Statistical properties of submm and FIR
({\it Herschel}) sources in Vela C.
          Median, average, maximum, and minimum values are given for each 
          physical parameter. Data for the 
{\it Herschel} sources are from Giannini et
           al.\ \cite{gianni12}.}             
\label{core:prop}      
\footnotesize
\centering                          
\begin{tabular}{c | c c c | c c c | c c c}        
\hline
    Parameter & \multicolumn{3}{c}{
{\it Herschel}} & 
      \multicolumn{3}{c}{LABOCA+
{\it Herschel}} & \multicolumn{3}{c}{LABOCA constant $T_{\rm dust}$}  \\
     & \multicolumn{3}{c}{Starless (218)} & \multicolumn{3}{c}{Starless (73)} & 
       \multicolumn{3}{c}{Starless (316)}  \\
     & median & average & min-max & median & average & min-max & median & average & min-max \\
\hline
$M(M_\sun)$ & $3.3$ & $5.5$ & $0.13-55.8$ & 
              $6.6$ & $9.9$ & $1.5-43.6$ & 
              $3.9$ & $6.1$ & $0.8-127.9$ \\
$T_{\rm dust}({\rm K})$ & $10.0$ & $10.3$ & $8.0-15.2$ & 
              $9.5$ & $9.6$ & $8.0-12.7$ & 
              $12.5$ & -- & -- \\ 
$D({\rm pc})$ & $0.064$ & $0.067$ & $0.025-0.13$ & 
              $0.080$ & $0.084$ & $(< D_{\rm beam})-0.151$ & 
              $0.078$ & $0.082$ & $(< D_{\rm beam})-0.151$ \\
\hline
          & \multicolumn{3}{c}{Protostellar (48)} & \multicolumn{3}{c}{Protostellar (124)} &
            \multicolumn{3}{c}{Protostellar (233)} \\ 
      & median & average & min-max & median & average & min-max & median & average & min-max \\
\hline
$M(M_\sun)$ &  $2.7$ & $4.8$ & $0.15-29.1$ & 
               $8.2$ & $14.4$ & $0.6-124.2$ & 
               $8.0$ & $16.3$ & $0.7-289.7$ \\
$T_{\rm dust}(K)$ & $11.4$ & $12.8$ & $9.0-24.2$ & $14.2$ & $15.0$ & $8.4-41.0$ & $12.5$ & -- & -- \\
$D({\rm pc})$ & $0.040$ & $0.040$ & $0.025-0.07$ & $0.091$ & $0.095$ & $0.040-0.155$ &
                $0.093$ & $0.094$ & $0.032-0.162$ \\
\hline
\end{tabular}
\end{table*}

By comparing the median mass of the two submm starless subsamples
(i.e. submm plus 
{\it Herschel} and submm with constant temperature), we note that this
roughly decreases to our estimated completeness limit when we use the whole starless sample
assuming a constant $T_{\rm dust}$. This indicates that many submm starless 
sources without an FIR-derived
temperature lie below the completeness limit.

\begin{figure}
   \centering
   \includegraphics[angle=-90,width=8cm]{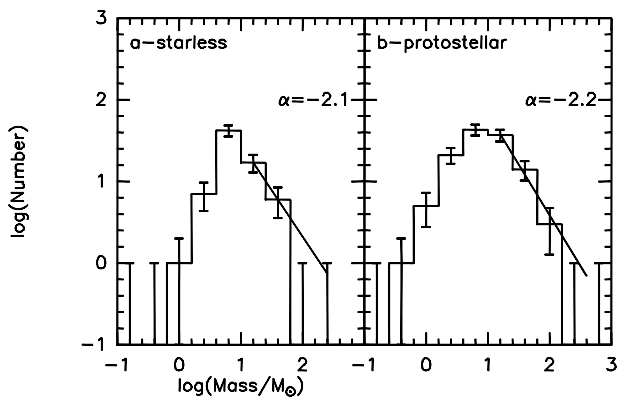}
   \caption{Core mass function for the submm starless sources (left) and
     the submm protostellar sources (right) whose temperature was
     derived from 
     {\it Herschel} data. Also shown, linear fits
     to the CMFs corresponding to a relation ${\rm d}N/{\rm d}M \sim M^{-2.1}$
     (starless cores) and ${\rm d}N/{\rm d}M \sim M^{-2.2}$
     (protostellar cores), respectively.
                      }
              \label{mass:hersch}%
    \end{figure}

\begin{figure*}
   \centering
   \includegraphics[angle=-90,width=16cm]{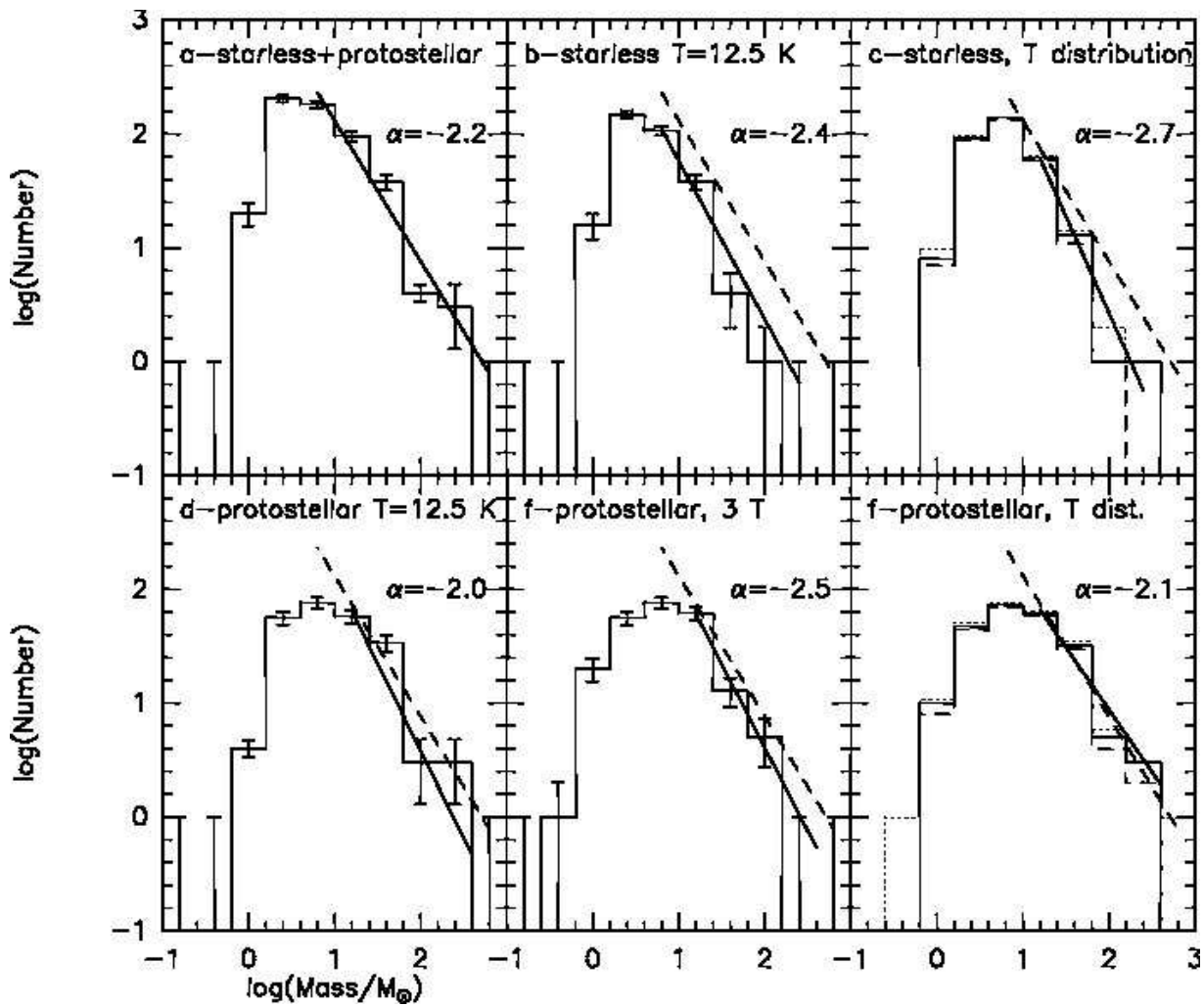}
   \caption{Comparison between core mass functions from the CuTEx 
     starless and protostellar source samples based on different 
     choices of dust temperatures. Solid straight lines are linear
     fits to the high-mass ends, and dashed straight lines show the slope found
     from the same fit to the total sample (starless plus protostellar).
     {\bf a} Same as in Fig.~\ref{cTdust:mass:hist};
     {\bf b} Starless cores with a constant dust temperature of $12.5$ K;
     {\bf c} Starless cores with temperature assigned based on
     a Monte Carlo procedure (see text);
     {\bf d} Protostellar cores with a constant dust temperature of $12.5$ K;
     {\bf e} Protostellar cores with three different temperatures
          assigned based on the core location (see text);
     {\bf f} Protostellar cores with temperature assigned based on
     a Monte Carlo procedure (see text). In panels {\rm \textbf{c}} and {\rm \textbf{f}}, the
     solid line histogram is the median (second quartile) distribution,
     the long-dashed line histogram is the first quartile distribution and the
     short-dashed line the third quartile distribution. In the other panels,
     errorbars indicate the Poissonian standard deviations of counts in each bin.
     Also indicated in each panel is the $\alpha$ index resulting
    from the linear fits, where ${\rm d}N/{\rm d}M \sim M^{\alpha}$.    
                      }
              \label{6pan:fig}%
    \end{figure*}

In addition, we performed a series of comparisons
between different instances of CMFs, all of which assume that they are linear in the log-log
space above the mass completeness limit. In this respect, we exploit two
different ways of computing the CMF slope $\alpha$: by a simple linear fit on a histogram,
and by using an unbiased
statistical estimator, namely the maximum likelihood (ML) estimator
(Maschberger \& Kroupa \cite{mushy}). 
The former makes for an easier interpretation, but the
latter is less prone to statistical bias. The results are listed in Table~\ref{fit:ml}.
The CMFs obtained for the submm starless and protostellar
cores with dust temperature derived from {\it Herschel} data 
are plotted in Fig.~\ref{mass:hersch}. The linear fit to the high-mass end
yields $\alpha = -2.1 \pm 0.4$ for the starless sample and $\alpha = -2.2 \pm 0.2$
for the protostellar sample. On the other hand,
the CMF of starless and protostellar cores derived for a constant
temperature of $12.5$ K are shown in Figs.~\ref{6pan:fig}(b) and (d).
The linear fit yields $\alpha = -2.4 \pm 0.2$ (starless cores) and $\alpha = -2.0 \pm 0.1$
(protostellar cores). The most remarkable difference between assuming a constant temperature and
using 
{\it Herschel} dust temperatures consists in the slightly
steeper $\alpha$ of the starless constant-temperature sample, 
although both slopes are consistent within 1 $\sigma$. 
However, the ML estimator results in an even steeper starless CMF ($\alpha = -2.8 \pm 0.2$), 
above a $2 \sigma$ significance level.
By comparing Figs.~\ref{6pan:fig}b and~\ref{mass:hersch}a
one can note that the single-core temperature approach (i.e. matching of 
{\it Herschel} and LABOCA sources) misses a fraction of objects
around the peak of the single-temperature CMF.  

To further test the effects of temperature, we used the so-called {\em bootstrap}
method adopting the temperature distributions 
of starless and protostellar 
{\it Herschel} cores obtained by Giannini et al.\ \cite{gianni12}. Thus, we
developed a simple Monte Carlo procedure to assign
a temperature drawn from these latter distributions to each source of our starless and protostellar samples. We constructed 1000 realisations each of the two CMFs.
The first three quartiles of each bin are shown in Figs.~\ref{6pan:fig}(c) and (f). A linear fit to the
higher-mass end of the median distributions (second quartile, which conserves the total number of sources in both cases)
yields $\alpha = -2.7 \pm 0.1$ (starless cores) and $\alpha = -2.1 \pm 0.1$
(protostellar cores), consistent with those derived from Figs.~\ref{6pan:fig}(b) and (d). 
In particular, this procedure enhances the higher steepness of the starless CMF.
The distributions corresponding to
the first and third quartiles show that 
the error per bin expected from temperature core-to-core
variations is similar to that due to the
Poissonian count statistics. 

%


   \begin{figure*}
   \centering
   \includegraphics[angle=-90,width=16cm]{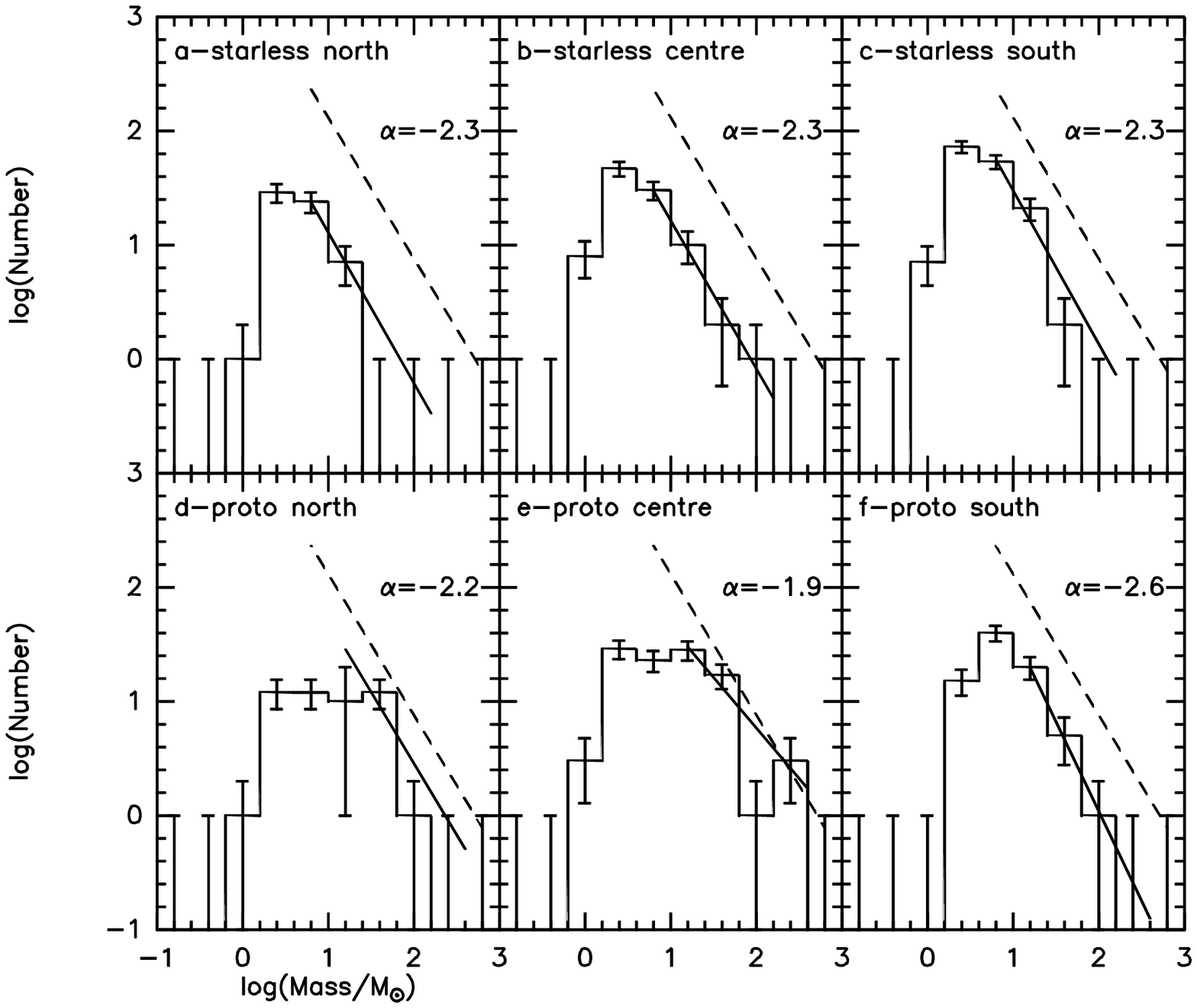}
   \caption{Comparison between core mass functions from the CuTEx starless and protostellar
     samples obtained in different parts of the cloud with a constant temperature of $12.5$ K.
     Solid straight lines are linear
     fits to the high-mass ends, dashed straight lines show the slope found
     from the same fit to the total sample (starless plus protostellar;
     see Fig.~\protect\ref{6pan:fig}a).
     {\bf a} Only starless sources in the northern part (DEC $> -43\degr 10\arcmin$); 
     {\bf b} Only starless sources in the central part ($-44{\degr} 15\arcmin <$ DEC $< -43{\degr} 10\arcmin$); 
     {\bf c} Only starless sources in the southern part (DEC $< -44{\degr} 15\arcmin$); 
     {\bf d} Only protostellar sources in the northern part (DEC $> -43\degr 10\arcmin$); 
     {\bf e} Only protostellar sources in the central part ($-44{\degr} 15\arcmin <$ DEC $< -43{\degr} 10\arcmin$); 
     {\bf f} Only protostellar sources in the southern part (DEC $< -44{\degr} 15\arcmin$). 
     Error bars indicate the Poissonian standard deviations of counts in each bin.
     Also indicated in each panel is the $\alpha$ index resulting
    from the linear fits, where ${\rm d}N/{\rm d}M \sim M^{\alpha}$.
                      }
              \label{3pan:fig}%
    \end{figure*}
%


We conclude that as long as the dust temperatures are randomly distributed among the cores,
this does not add further uncertainty to the CMF compared to that expected from the Poissonian
statistics. A similar result was obtained by Csengeri et al.\ \cite{csen2} on a smaller
sample of clumps using Monte Carlo methods.  
However, {systematic} differences in temperatures could still significantly  affect
the CMFs. Based on the temperature distributions shown in Fig.~\ref{histo:temp}, 
systematic temperature differences may be expected to  significantly affect only the protostellar
CMF, since the spread in temperature is relatively narrow for the starless subsample. 
The only systematic difference that can be envisaged is a temperature gradient from
south to north (e.g. Fissel et al.\ \cite{fissel}). 
To test the effects, we derived the CMF of protostellar
cores by assigning a constant temperature of $12.5$ K for DEC $< -44{\degr}$,
20 K for $-44{\degr} <$ DEC $< -43{\degr} 30\arcmin$,
and 16 K for DEC $> -43{\degr} 30\arcmin$. The results are shown in
Fig.~\ref{6pan:fig}e. A linear fit to the high-mass end of the protostellar CMF yields $\alpha =
-2.5 \pm 0.2$, steeper than all other determinations of the protostellar CMF
(see Figs.~\ref{6pan:fig}(d) and (f) and Table~\ref{fit:ml}). On the other hand, the ML estimator
yields $\alpha = -2.2 \pm 0.2$, suggesting no significant differences. 

We also wanted to investigate spatial changes in the CMF throughout Vela C. Therefore,
we constructed the CMFs for a constant dust temperature of $12.5$ K
by selecting sources in the three areas defined in Sect.~\ref{cotemp} (north, centre, and south).
We note that the number of sources are comparable in the centre and south areas 
(104 starless and 95 protostellar cores in the centre, 
152 starless and 81 protostellar cores in the south), but are roughly half as much in the
north (69 starless and 48 protostellar cores). 
The results are shown in
Fig.~\ref{3pan:fig}. As for the starless samples, their CMFs
do not exhibit any significant variations above a $1 \sigma$ level with position. 
The slopes of the high-mass end of the protostellar CMFs 
are flatter in the north and centre than in the south. The ML estimator confirms this trend
at a higher confidence level. We also note that any systematic
difference in the temperature of protostellar cores between the three areas, as tested above, 
would only shift the CMFs in log(mass), but would not affect the CMF shapes and slopes. 

One concern is that the protostellar and starless CMFs in the centre may be affected
by the saturation in the WISE band and the use of MSX towards RCW~36. A look at the
submm cores in this area shows that only four starless cores are above the mass
completeness limits, three of which have masses (in the single-temperature approximation) of 
128, 44, and 39 $M_{\sun}$, respectively. The only effect that one can envisage is that some or all of these
are actually protostellar in nature. This would at most result in a further steepening of the
starless CMF and a further flattening of the protostellar CMF in the centre.
We also assessed how the poorer statistics in the three areas may bias the results by performing 
new fits after shifting the (north, centre, south) 
CMF bins by $0.2$ (in logarithm). We obtained CMFs that are slightly flatter
in all cases (the point at $\log(M/M_{\sun})=0.6$ was included in the linear fits), but
consistent with the previous determinations at a $1 \sigma$ level.

\begin{table*}
\caption{Spectral indices $\alpha$ (with ${\rm d}N/{\rm d}M \sim M^{\alpha}$)
        from linear fits and the ML estimator. 
        To minimise incompleteness effects,
        the adopted lower-mass ends are $\log(M/M_{\sun}) = 0.8 - 1.2$
        for the fits (i.e. the central value of the bin next to the peak)
         and $\log(M/M_{\sun}) = 0.8$ for the ML computations
        ($1.2$ and $1.2$, respectively, for CLUMPFIND).
         }             
\label{fit:ml}      
\centering                          
\begin{tabular}{l c c c c c c c }        
\hline\hline                 
    &             & \multicolumn{2}{c}{protostellar+starless} & 
                                \multicolumn{2}{c}{protostellar} & \multicolumn{2}{c}{starless}\\ 
CMF & temperature & linear fit & ML & linear fit & ML & linear fit & ML \\    
    &     (K)     &            &    & & & & \\
\hline                        
  CuTEx & $12.5$ & $-2.2 \pm 0.2$ & $-2.2 \pm 0.1$ & 
                                      $-2.0 \pm 0.1$ & $-1.9 \pm 0.1$ & $-2.4 \pm 0.2$ & $-2.8 \pm 0.2$ \\
  CLUMPFIND & $12.5$ & $-2.1 \pm 0.2$ & $-2.0 \pm 0.1$ & -- & -- & -- & -- \\
  CuTEx & 
{\it Herschel} match & -- & -- & $-2.2 \pm 0.2$ & $-1.9 \pm 0.1$ & $-2.1 \pm 0.4$ & $-2.0 \pm 0.2$ \\
  CuTEx & Monte Carlo & -- & -- & $-2.1 \pm 0.1$ & -- & $-2.7 \pm 0.1$ & -- \\
  CuTEx & 3T\tablefootmark{a} & -- & -- & $-2.5 \pm 0.2$ & $-2.2 \pm 0.2$ & -- & -- \\ 
  CuTEx (north) & $12.5$ & -- & -- & $-2.2 \pm 0.5$ & $-1.5 \pm 0.1$ & $-2.3 \pm 0.4$ & $-2.1 \pm 0.3$\\
  CuTEx (centre) & $12.5$ & -- & -- & $-1.9 \pm 0.2$ & $-1.7 \pm 0.1$ & $-2.3 \pm 0.2$ & $-2.2 \pm 0.3$ \\
  CuTEx (south) & $12.5$ & -- & -- & $-2.6 \pm 0.3$ & $-2.5 \pm 0.2$ & $-2.3 \pm 0.2$ & $-2.7 \pm 0.3$ \\
\hline                                   
\end{tabular}
\tablefoot{
\tablefoottext{a}{Three different dust temperatures were used: $12.5$ K for
  cores in the south, 20 K for cores in the centre, and 16 K
  for cores in the north of Vela C (see text).}
}
\end{table*}
%

\subsection{Clustering and large-scale structures}
\label{clu:ls}

\begin{figure}
   \centering
   \includegraphics[angle=-90,width=8cm]{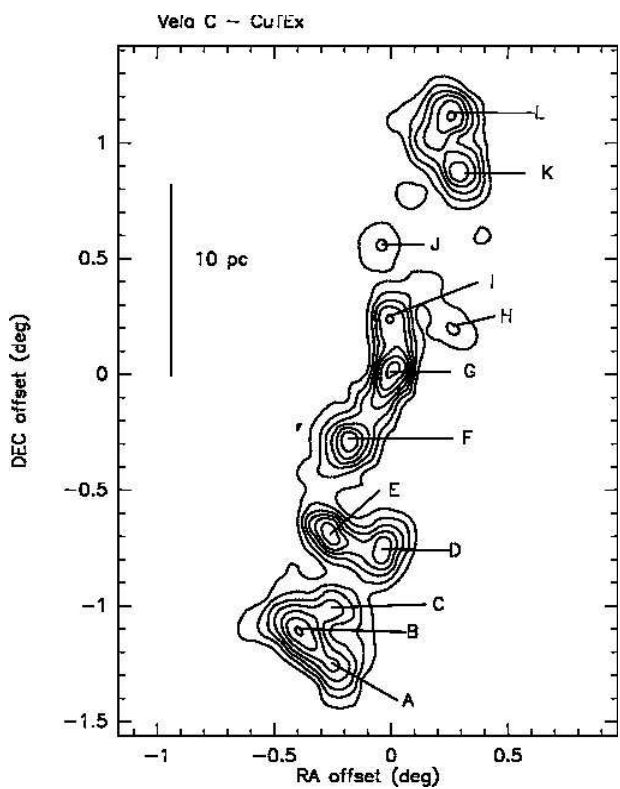}
   \caption{Surface density map of CuTEx sources down to $0.3$ Jy
      of flux density. Contours range from 100 sources deg$^{-2}$ 
        in steps of 150 sources deg$^{-2}$. Coordinate offsets are
        with respect to the position of the ionisation source
        of RCW~36 (RA$=08^{\rm h}59^{\rm m}27^{\rm s}$, DEC$=-43\degr 45\arcmin
        26\arcsec$). 
                      }
              \label{dens:map}%
    \end{figure}

To probe core clustering in Vela C, we counted all CuTEx sources down to a
flux density of $0.3$ Jy (the estimated completeness limit)
inside squares of side $\sim 9\arcmin$ 
($\sim 1.8$ pc at 700 pc) parallel to the 
right ascension axis, offset by half the square side (i.e. a Nyquist
sampling interval) in both right ascension and declination. The resulting map of surface density
is shown in Fig.~\ref{dens:map}. Clearly, at least 12 clusters of cores can
be retrieved; these clusters are almost equally spaced and have sizes of $\sim 2$ pc (therefore unresolved). This probably 
reflects large-scale fragmentation of the pristine molecular cloud. 
Positions and peak source surface densities of the clusters 
are listed in Table~\ref{dens:tab}.

%

\begin{figure}
   \centering
   \includegraphics[angle=-90,width=8cm]{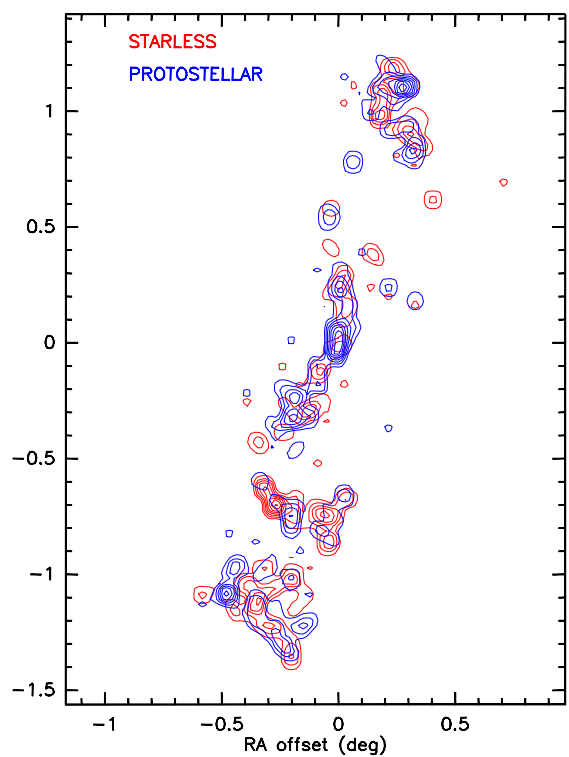}
   \caption{As in Fig.~\protect\ref{dens:map}, but for starless (red contours) and
    protostellar (blue contours) cores separately. Only the highest
    contours are plotted (from 200 sources deg$^{-2}$
        in steps of 200 sources deg$^{-2}$). Coordinate offsets are
        with respect to the position of the ionisation source
        of RCW~36 (RA$=08^{\rm h}59^{\rm m}27^{\rm s}$, DEC$=-43\degr 45\arcmin
        26\arcsec$).
                      }
              \label{surf:pre:pro}%
    \end{figure}

A look at Fig.~\ref{dist:pre:proto} suggests that protostellar cores may
concentrate towards the filaments. By repeating the same procedure as
above for starless and protostellar cores separately (after decreasing the
square size to $4\farcm 5$) we could not find any significant 
morphological difference in the lowest density contours. Nevertheless, starless and
protostellar projected distributions appear to peak at slightly different locations 
(see Fig.~\ref{surf:pre:pro}). This hints at least at a partial segregation of regions hosting
starless and protostellar cores. Furthermore, a comparison of the submm and FIR 
maps suggests a population of starless, low-mass dust cores (below our submm completeness
limit) outside the cloud filaments (see Sect.~\ref{cotemp}) that should be further
investigated with more sensitive submm observations. 

%


\begin{table}
\caption{Position (offsets from RA$=08^{\rm h}59^{\rm m}27^{\rm s}$, 
         DEC$=-43\degr 45\arcmin 26\arcsec$) and peak surface density of the clusters of
         submm cores
         in Vela C.}             
\label{dens:tab}      
\centering                          
\begin{tabular}{c c c c}        
\hline\hline                 
Cluster & RA offset & DEC offset & Peak surf. density \\    
  ID  &   (deg)  &   (deg)         &   (pc$^{-2}$) \\
\hline                        
 A & -0.25 & -1.26 & 5 \\ 
 B & -0.40 & -1.11 & 7 \\ 
 C & -0.25 & -1.02 & 4 \\ 
 D & -0.03 & -0.76 & 5 \\ 
 E & -0.26 & -0.69 & 6 \\ 
 F & -0.18 & -0.29 & 7 \\ 
 G & 0.01 & 0.03 & 7 \\ 
 H & 0.27 & 0.20 & 2 \\ 
 I & 0.00 & 0.24 & 5 \\ 
 J & -0.04 & 0.56 & 2 \\ 
 K & 0.29 & 0.88 & 5 \\ 
 L & 0.26 & 1.12 & 6 \\ 
\hline                                   
\end{tabular}
\end{table}
%

\section{Discussion}
\label{dissec}
\subsection{Comparison with other studies of the Vela Molecular Ridge}
 
The submm CMFs can easily be compared 
with those derived for Vela C in other works. 
Netterfield et al.\
\cite{Nett09} derived power-law indices $\alpha = -1.95 \pm 0.05$ for
sources with $T_{\rm dust} > 14$ K and $\alpha = -3.22 \pm 0.14$
for sources with $T_{\rm dust} < 14$ K  from their 
BLAST FIR observations of the whole Vela region. The first value must be
compared with that derived from the linear fits to our protostellar sample 
and in fact is
fully consistent with it (see Table~\ref{fit:ml}). On the other hand, the second value
is not consistent with our linear fits to the starless sample.
Nevertheless, we note that Netterfield et al.\
\cite{Nett09} derived  $\alpha = -2.9  \pm 0.2$ for cold cores
in Vela C, which is closer to our linear fit ($\alpha = -2.4  \pm 0.2$,
see Table~\ref{fit:ml}) and quite consistent 
with the ML value ($\alpha = -2.8  \pm 0.2$) for starless cores
in single-temperature approximation.

Giannini et al.\ \cite{gianni12} found $\alpha = -2.1  \pm 0.2$
for their sample of {\it Herschel} prestellar cores in Vela C,
which is fully consistent with our linear fit to the subsample of
starless cores with single-core temperatures ($\alpha = -2.1  \pm 0.4$,
see Table~\ref{fit:ml}). These authors note that by limiting their
fit to the CMF at $M > 14$ $M_{\sun}$, the completeness limit
of BLAST data, one obtains $\alpha = -2.9  \pm 0.2$, which is consistent
with our ML value for the starless sample in single-temperature approximation.
This would confirm that the {\it Herschel} sample is missing a significant fraction of
low-mass ($M < 14$ $M_{\sun}$) cores compared to the submm sample, 
which would explain the discrepancy
between the CMFs of starless cores in the single-temperature approximation  
and the single-core temperatures (i.e. from the associated {\it Herschel}
cores) already noted ($\alpha \sim -2.4, -2.8$ vs.\ $\alpha \sim -2.1$);
although this difference is only marginally significant.

As for other clouds in the Vela Molecular Ridge,
Massi et al.\ \cite{massi07} found $\alpha = -1.45$ using
CLUMPFIND on their mm map (250 GHz) of Vela D. As they do not
differentiate between starless and protostellar cores, this value
must be compared to the value that we obtained from the whole sample
using the same algorithm, that is,  $\alpha = -2.1 \pm 0.2$ 
(see Table~\ref{fit:ml}). They claim a mass completeness limit of 1--$1.3$ $M_{\sun}$
and their most massive core is $\sim 100$ $M_{\sun}$, less than found in 
Vela C with CLUMPFIND. However, Olmi et al.\ \cite{holmes}
revised the power-law index for Vela D to $\alpha = -2.3 \pm 0.2$ by 
combining the data
of Massi et al.\ \cite{massi07} and BLAST FIR photometry. They
concluded that the value derived by Massi et al.\ \cite{massi07} is
likely to be biased due to incompleteness. The mass spectra of dense
cores in the two clouds are therefore consistent with each other, as far as
protostellar and starless cores are considered together. 

We compared the mass of a few submm cores with those derived by
Minier et al.\ \cite{minier} for seven clumps identified in
the same area (towards RCW~36).
We found some discrepancies that are discussed in the online
Appendix (Appendix~\ref{comp:app}) and are
probably due to the different measuring techniques and clump/core definition
adopted.

\begin{figure}
   \centering
   \includegraphics[angle=-90,width=8cm]{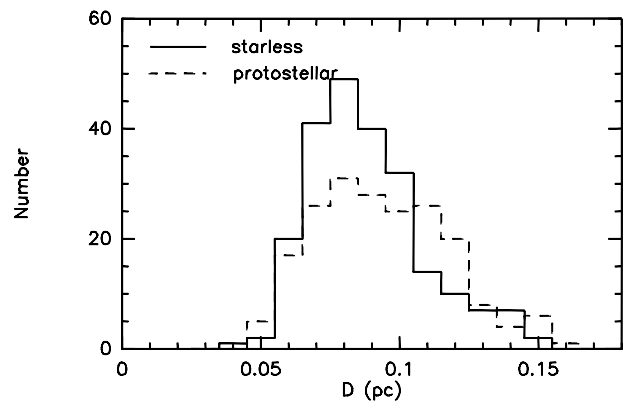}
   \caption{Deconvolved size (diameter) distribution
     for starless (solid line) and protostellar (dashed line) submm cores.
     Only sources with minimum deconvolved size greater than half the beam FWHM have
    been taken into account.
                      }
              \label{dec:siz}%
    \end{figure}

Other aspects of our results can be explored through a comparison with
the sample of IR cores of Giannini et al.\ \cite{gianni12}. 
The most remarkable difference between the {\it Herschel} FIR and LABOCA submm
statistics (see Table~\ref{core:prop})
is that the sources in the protostellar sample are on average more massive than the 
starless sources, as far as submm cores are concerned. In addition, the FIR sources
are on average smaller than the submm ones, which is clearly due to using the more 
resolved 160 $\mu$m map to set the {\it Herschel} source size. 
The deconvolved sizes of both starless and submm protostellar sources are plotted in
Fig.~\ref{dec:siz}. This can be compared with Fig.~4 of
Giannini et al.\ \cite{gianni12}. The distribution of the submm starless cores 
resembles that of the {\it Herschel} starless cores, peaking at $\sim 0.08$ pc 
(compared to $0.05$ pc) and declining at $D > 0.1$ pc. 
On the other hand, protostellar submm cores display a size distribution
that is different from that of both {\it Herschel} starless and 
{\it Herschel} protostellar cores. While 
Giannini et al.\ \cite{gianni12} found that the protostellar sources are quite
compact ($0.04$ to $0.07$ pc), in our submm sample protostellar cores are even less 
compact than starless ones.  

%


\subsection{Core properties}

As can bee seen from Table~\ref{fit:ml}, most of our CMF determinations
are in very good agreement
with a Salpeter IMF ($\alpha = -2.35$). 
This should be considered as a robust result,
based on our discussion above. In fact, the ML estimator suggests a slightly flatter CMF for
protostellar cores and a slightly steeper one for starless cores. 
As pointed out in Sect.~\ref{s:p:c}, the starless CMF is
actually representative of a real prestellar core population, so it can be compared
with prestellar CMFs derived in other regions. 
Table~\ref{near:sftab} lists the results of a number of surveys carried out towards nearby star-forming regions. 
Prestellar CMFs of nearby regions (125 to 700 pc) obtained from single-dish observations
with spatial resolutions in the range $0.01 - 0.07$ pc, though sometimes shifted to higher 
masses, are all consistent with a
Salpeter IMF in their high-mass end.
Olmi et al.\ \cite{olmi} found that this similarity persists at a larger scale, deriving
an average slope $\alpha = -2.12 \pm 0.15$ in a sample of {\it clump} mass functions
extracted from the Hi-Gal catalogue.
We note that in the most recent prescriptions the standard IMF 
is steeper than a Salpeter one; for example, Kroupa et al.\ \cite{kroupa}
propose $\alpha = -2.7$ for $M \geq 1 M_{\sun}$ whereas Scalo \cite{scalo}
find $\alpha = -2.7 \pm 0.5$ in the range 1--10 $M_{\sun}$. This agrees
with the spectral index we found using the ML estimator for starless cores.  

%
%
\begin{table*}
         \caption{CMF determinations in nearby star-forming regions }             
\label{near:sftab}      
\footnotesize
\centering                          
\begin{tabular}{c | c | c | c | c }        
\hline
Region & Distance & Beam Size & CMF\tablefootmark{a} & References \\
        & (pc)    &           &     &           \\
\hline
Pipe Nebula & 160 & $\sim 0.03$ pc & shifted IMF & Alves et al.\ \protect\cite{alves} \\
Orion regions & $\sim 400$ & $14\arcsec$ at 850 $\mu$m & shifted IMF & Nutter \& Ward-Thompson \protect\cite{nutter} \\
\object{Polaris Flare} & 150 & $18\arcsec$ at 250 $\mu$m & shifted IMF & Andr\'{e} et al.\ \protect\cite{andrew} \\
\object{Aquila Rift} & 260 & $18\arcsec$ at 250 $\mu$m & $\alpha = -2.33 \pm 0.06$ & Konyves et al.\ \protect\cite{kony} \\
\object{Orion A} & $\sim 400$ & $18\arcsec$ & $\alpha \sim -2.3$ & Ikeda et al.\ \protect\cite{ikeda} \\
Perseus, Serpens, Ophiuchus & 125 -- 260 & $31\arcsec$ &  $\alpha = -2.3 \pm 0.4$ & Enoch et al.\ \protect\cite{enoch} \\
Taurus \object{L1495} cloud\tablefootmark{b} & 140 & $\sim 18\arcsec$ & IMF & Marsh et al.\ \protect\cite{marsh} \\
\hline
\end{tabular}
\tablefoot{
\tablefoottext{a}{{\it Shifted IMF} means that the whole CMF approximates a standard IMF when shifted to lower masses.} 
\tablefoottext{b}{Sources retrieved using the {\it getsources} algorithm of
   Men’shchikov et al.\ \protect\cite{men}}
    }
\end{table*}

ALMA is now starting to address farther regions ($d > 2$ kpc) but with very good angular 
resolution. This is instrumental in probing different environments such as infrared
dark clouds (IRDCs) and massive star-forming regions. 
Some prestellar CMFs recently derived from ALMA data are listed in Table~\ref{far:sftab}.
We note the cases for variations in the high-mass end of the CMF 
reported by Motte et al.\ \cite{motte17} and Liu et al.\ \cite{mengyao}.
Both find top-heavy CMFs. 
However, these ALMA
data are not easy to compare with single-dish data of nearby regions. As discussed by
Liu et al.\ \cite{mengyao}, i) prestellar and protostellar cores cannot be
separated given the larger distances, ii) lack of spatial frequencies has to be taken into
account, and iii) the statistics are often poor. In addition, the statistics
show a higher sensitivity to the adopted core-retrieval algorithms.  

%
%
\begin{table*}
         \caption{CMF determinations in farther star-forming regions }             
\label{far:sftab}      
\footnotesize
\centering                          
\begin{tabular}{c | c | c | c | c | c | c}        
\hline
Region & Distance & Beam Size & CMF & Mass range & Algorithm\tablefootmark{a} & References \\
        & (kpc)    &           &     &  ($M_{\sun}$) &   &         \\
\hline
\object{IRDC G14.225--0.506} & $\sim 2$ & $\sim 3\arcsec$ & $\alpha = -2.6 \pm 0.7$ & $2.4$--14 & CLUMPFIND & Ohashi et al. \protect\cite{ohashi}\\
\object{G286.21+0.17} & $\sim 2.5$ & $\sim 1 \arcsec$ & $\alpha = -2.24 \pm 0.17$ & $\sog 1$ & dendrogram & Cheng et al.\ \protect\cite{cheng}\\
G286.21+0.17 & $\sim 2.5$  & $\sim 1 \arcsec$ & $\alpha = -1.64 \pm 0.13$ & $\sog 1$ & CLUMPFIND & Cheng et al.\ \protect\cite{cheng}\\
\object{W43--MM1} & $\sim 5.5$ & $\sim 0\farcs44$ & $\alpha = -1.90 \pm 0.06,-1.96 \pm 0.02$ & $1.6$ -- 100 & getsources & Motte et al.\
\protect\cite{motte17}\\
7 IRDCs & $2.4 - 5.7$ & $\sim 1 \arcsec$ & $\alpha = -1.70 \pm 0.13$ & $\ge 1.26$ & dendrogram & Liu et al.\ \protect\cite{mengyao}\\
\hline
\end{tabular}
\tablefoot{
\tablefoottext{a}{References for core retrieving algorithms. CLUMPFIND: Williams et al.\
    \protect\cite{willy}; dendrogram: Rosolowsky et al.\ \protect\cite{rosolo};
    getsources: Men’shchikov et al.\ \protect\cite{men}}
    }
\end{table*}

It is useful to study other global properties of the submm cores such as the mass--size relation.
We derived a mass--size relation $M \sim D^{2.2}$ (see Sect.~\ref{statcores})
for the whole (protostellar and starless) sample. Considering that
the power-law index is sensitive to the core-retrieval algorithm ($M \sim D^{1.7}$
when using the CLUMPFIND output) and cautioning against possible incompleteness effects,
this is consistent with the mass--radius relation for clumps and molecular clouds
$M \sim R^{1.9}$ found by Larson \cite{larson}. 
For the sake of comparison, the results of a few large-scale surveys are listed in 
Table~\ref{mass:sizetab}; all these studies suggest
that molecular cores and clumps arrange their masses 
according to $M \propto D^{2}$, which is expected for ensembles
of structures with the same column density over a large range of sizes.   

%
%
\begin{table}
         \caption{Mass--size relations ($M \sim D^{X}$) from the literature}             
\label{mass:sizetab}      
\scriptsize
\centering                          
\begin{tabular}{l | l | l | l }        
\hline
Sample & $X$ & References & Notes \\
\hline
Clumps and molecular clouds & $1.9$ & Larson \protect\cite{larson} & \\
Clumps and cores in Vela D & $1.74$ & Massi et al.\ \cite{massi07} & (1) \\
Clumps and cores in Taurus & $\sim 1.9$ & Marsh et al.\ \cite{marsh} \\
ATLASGAL sample of clumps &  &  &  \\
associated with methanol masers, & & & \\
YMOs, and HII regions & $1.67$ & Urquhart et al.\ \cite{james}  & (2) \\
Galactic clumps from MALT90 & $2.2$ & Contreras et al.\ \cite{contreras} & (3) \\
\hline
\end{tabular}
\tablefoot{(1) CLUMPFIND, adopting constant $T_{\rm dust} = 30$ K; (2) adopting constant $T_{\rm dust} = 20$ K;
(3) $T_{\rm dust}$ and column densities from {\it Herschel}
and ATLASGAL data.
    }
\end{table}

A mass--size sequence, irrespective of
the value of its power-law index, has important implications 
concerning the evolution of dense prestellar cores. Either these cores are quickly
assembled, or they evolve inside the mass--size sequence itself.
In the latter case, two scenarios can be envisaged. Simpson et al.\
\cite{simpson} suggested that cores are thermally supported 
and move along the mass--size sequence from the low-mass
end, by accreting mass quasi-statically while maintaining a Bonnor-Ebert equilibrium. 
Once they reach Jeans instability, they collapse as protostars.   
Since starless cores in Vela C lie well above the mass instability limits, this model
should be modified by introducing other forms of inner physical support (magnetic fields, 
turbulence). 
However, we note that the detection loci drawn in Fig.~\ref{mass:size:fig} 
would hide any signature of evolutionary paths from the critical Bonnor-Ebert half-mass
loci to the mass--size sequence. In addition, the adopted dust opacity affects
the picture. In fact, using the ATLASGAL value would move the sequence and
detection loci nearer to the Bonnor-Ebert loci. 
Nevertheless, studies of magnetic fields in Vela C 
based on FIR polarisation (Fissel et al.\ \cite{fissel18}) suggest a connection
between large-scale magnetic fields and small gas structures, which deserves
further investigation. 
Alternatively, the fact that a similar mass--size relation is found
for clumps and cores may indicate that dense cores move down the sequence from the high-mass end by fragmentation (e.g. Elmegreen \& Falgarone \cite{elfal}). 

Hydrodynamic simulations can help us to understand dense core evolution as well. 
Li et al.\ \cite{li} performed turbulent simulations of a magnetised cloud,
deriving both the global physical properties and the evolution of bound
dense cores. Their CMF is consistent at the high-mass end with a power law
of index $\alpha \sim -2.29$, but the mass--size relation is steeper
than observed in Vela C ($M \sim D^{2.7}$). Interestingly, the CMF flattens
with cloud age. 
Offner et al.\ \cite{offner} simulated both driven and decaying turbulence
in molecular clouds. They obtained bound cores exhibiting a CMF with a
high-mass end consistent with a stellar IMF but a mass--size relation flatter
than ours ($M \sim D$). However, their statistics are poor and the core sizes they obtain
are $< 0.02$ pc, less than our adopted spatial size limit. 
Offner \& Krumholz \cite{offner:krum} also analysed the shapes of cores from
simulations with driven and decaying turbulence. The axis ratio distributions
they obtained roughly resemble ours, with median and mean values in the range
$0.6$--$0.7$ and minimum values of  $\sim 0.2$. Their size distributions cannot be
compared with ours, which is biased by having excluded cores
smaller than half the LABOCA beam size. However, prestellar and protostellar synthetic 
cores are smaller and distributed in a smaller range than those derived from
Orion observations by Nutter \& Ward-Thompson \cite{nutter}.  Prestellar core
formation has also been investigated by Gong \& Ostriker \cite{gong15} and Gong \&
Ostriker (\cite{gong11}, \cite{gong09}) for converging flows triggering star formation
in the post-shock gas. In particular, by retrieving dense cores in
snapshots of their simulations (which include turbulence and gravity with inflow
Mach numbers in the range 2--16),
Gong \& Ostriker \cite{gong15} can reproduce mass--size relations consistent with
ours (e.g. those with inflow Mach numbers 4 and 8).
Since their cores gain mass during their evolution, which occurs with lifetimes of
$\sim 1$ Myr, the observed mass--size sequence would enclose condensations in different stages
of evolution. Notably, their simulations also reproduce a CMF which is consistent with
a stellar IMF, although steeper in the high-mass end. 
By selecting all cores in the stage just before a sink particle is formed
(i.e. on the verge of forming a protostar) they obtain both
a shallower mass--size relation and a CMF with a steeper high-mass end, implying 
a significant deficit of high-mass cores. 

Lee et al.\ \cite{lee} applied the gravoturbulent fragmentation theory of Hennebelle \& Chabrier
to a filamentary environment with magnetic fields, 
finding CMFs that resemble a Salpeter IMF at their high-mass end. Interestingly, these authors 
found that filaments with high mass per length ($> 1000$ $M_{\sun}$pc$^{-1}$) are needed
to produce massive cores. Otherwise, the CMF is steeper than a Salpeter IMF.
Guszejnov \& Hopkins \cite{guz:hop} modelled the fragmentation of self-gravitating, 
turbulent molecular clouds from large to small scales semi-analytically, confirming 
that the CMF shape is similar to the IMF at the high-mass end. 

Undoubtedly, all simulations point to turbulence and magnetic fields as important
ingredients in shaping some of the statistical properties of prestellar cores in Vela C, 
such as CMF and the mass--size relation. However, our observation do not capture which mechanisms are the dominating ones.

\subsection{Star formation activity in Vela C}
\label{sfavc}

Vela C exhibits different levels of star formation activity in its various parts.
The five regions identified by Hill et al.\ \cite{hill11} roughly contain the same mass,
but the peak column density in the areas referred to as nests by Hill et al.\ \cite{hill11} and in the north is smaller than in
the ridges ($\sim 1$ vs.\ $\sim 2 \times 10^{23}$ cm$^{-2}$; Hill et al.\ \cite{hill11}).
A large star cluster with massive stars and an HII region (RCW~36), and an embedded young cluster
with intermediate-mass stars (IRS~31) have already formed in the centre nest and in the north, 
respectively, whereas the South-Nest is the coldest region of Vela C (Hill et al.\ 
\cite{hill11}). 

As shown in Sect.~\ref{s:p:c}, the fraction of protostellar cores varies along the cloud; it
is smaller in the south and larger in the centre. If taken at face value, this would agree
with a picture where the southern part of Vela C is the least evolved and the central part
(hosting RCW~36) the most evolved. However, possible biases have to be taken into account.
Baldeschi et al.\ \cite{baldeschi} studied how the physical parameters of detected compact 
sources are affected by degradation of spatial resolution. Using {\it Herschel} maps of
nearby star-forming regions, they simulated the effect of an increasing distance.  
They clearly found a significant decrease in the fraction of prestellar cores and a 
consequent increase in the fraction of protostellar cores up to 1 kpc. Thus, an
increasing number of prestellar cores is misidentified due to lack of spatial
resolution. When prestellar cores lie in projection very near to a protostellar core,
after degrading the spatial resolution they merge into one large protostellar core. A look
at Fig.~\ref{3pan:fig} shows that the most massive (and largest) cores were detected
in the centre of Vela C. This region is dominated by the Centre-Ridge, which hosts
RCW~36. Minier et al. \ \cite{minier} propose that RCW~36 formed in a sheet of gas
roughly edge-on and is driving an expanding bubble of matter
yielding an edge-on ring. This
would explain the large maximum column density derived by Hill et al.\ \cite{hill11}.
If so, a significant degree of line-of-sight confusion could arise, leading 
to larger cores and an increased fraction of protostellar cores due to the effect envisaged
by Baldeschi et al.\ \cite{baldeschi}. This could also explain the slightly flatter CMF
derived in the central (and northern) part of Vela C (see Fig.~\ref{3pan:fig}). Furthermore, the size
distribution of both starless and protostellar cores (see Fig.~\ref{dec:siz})
hints at a fraction of cores resulting from confusion of smaller condensations,
which could explain why protostellar
cores are less compact than starless ones. When a number of cores are merged into
a single larger one by the finding algorithm, there is a high probability that one of them
is a protostar, biasing the whole output as protostellar. Thus, larger cores are more likely to be
labelled as protostellar. In this
respect, it would be interesting to observe the largest cores with ALMA.

Notably, the CMF is slightly steeper in the southern part of Vela C both for starless
and protostellar cores, although these are consistent with each other within 1$\sigma$
(Fig.~\ref{3pan:fig} and Table~\ref{fit:ml}). There is clearly a shortage of high-mass
cores. This fact may further suggest that this is the least evolved area in Vela C, where
massive cores have not yet been formed. In turn, this might also indicate that low- and
intermediate-mass stars have to form earlier than massive stars.

Based on our data, we propose that Vela C is composed of sheets of
filaments with different orientations; we propose that nests are seen roughly face-on, and ridges
are seen almost edge-on. The South-Nest is likely to be
in an earlier phase of global contraction, whereas at least
RCW~36 in the Centre-Ridge may originate from converging flows,
explaining why this part of the cloud has already evolved
into a massive star-forming region. 
This scenario would be consistent with observational studies of magnetic field orientation
in the region (Kusune et al.\ \cite{takayoshi}, Soler et al.\ \cite{soler},
Fissel et al.\ \cite{fissel18}), pointing to magnetic field enhancement by
density increase in the Centre-Ridge (consistent with gas compression) and
showing that the plane-of-the-sky magnetic field direction is chaotic in the South-Nest. 
In addition,  Sano et al.\ \cite{sano} suggest that RCW~36 formed $\sim 0.1$ Myr ago due to
cloud-cloud collision.

In Sect.~\ref{clu:ls}, we describe our investigation of how dense cores cluster. When 
using a spatial resolution of $\sim 1.8$ pc, clusters of cores less than 2 pc
in size are found to be separated by projected distances in the range 2--6 pc. Multiple clusters of dense
cores are retrieved inside each of the sub-regions identified by Hill et al.\ \cite{hill11}
through a multiple-resolution analysis decomposition of Vela C at a scale of 2 pc
(see their Fig.~7). When using a
spatial resolution of $\sim 0.9$ pc and counting starless and protostellar cores
separately, partially segregated groups of starless and protostellar cores 
are retrieved with separations in the range 1--2 pc. These roughly correspond
to the high column density regions ($> 10^{22}$ cm$^{-2}$) identified by
Hill et al.\ \cite{hill11}. The core clustering is then a signature of gas fragmentation.
In fact, these scale-lengths are roughly consistent both with a magnetised and a turbulent
Jeans length for $B \sim 30$ $\mu$G or a velocity dispersion $\sim 2$ km s$^{-1}$
and a mean gas density $n_{H_2} = 10^{3}$ cm$^{-3}$. Fissel et al.\ \cite{fissel18}
suggest that the magnetic field in Vela C must be globally at least trans-Alfv\'{e}nic,
hence important in the evolution of cloud structures. A simple Jeans thermal analysis
is therefore not appropriate for Vela C. 
Hierarchical core clustering and gas fragmentation down
to scales of $\sim 0.1$ pc (i.e. less than the spatial resolution of our data)
have been found in other nearby star-forming regions
(e.g. Kainulainen et al.\ \cite{kainu}, Mattern et al.\ \cite{mattern}).
 
The star formation activity of Vela C can be qualitatively compared with that of Vela D
(Massi et al.\ \cite{massi06}, \cite{massi07}; Elia et al.\ \cite{elia07};
Olmi et al.\ \cite{holmes}). Here, young embedded star clusters of $\sim 1$ Myr in age
are located towards gas clumps inside filaments which outline the edges of expanding
bubbles. This suggests that the newly born stars are actually a second-generation 
population and therefore that Vela D is more evolved than Vela C. 
 
\section{Conclusions}
\label{consec}

We studied the dust emission in
a large-scale map of the GMC Vela C obtained at
345 GHz ($0.87$ mm) with LABOCA at the APEX submm telescope. Using the algorithm
CuTEx, we decomposed the emission into 549 compact submm sources
(plus another 15 associated with the HII region RCW~34), with sizes ranging from less than
half the APEX beam ($0.03$ pc) to $0.16$ pc. The algorithm CLUMPFIND was also used
for comparison purposes, retrieving 291 compact submm sources (plus another 104 with
beam-convolved angular sizes smaller than the APEX beam size). CuTEx is more focused on 
detecting compact sources, so we have based our analysis on its output only. Here, the submm sources
are referred to as dense cores, which is appropriate given the size range spanned. 
We used ancillary MIR (WISE)
and FIR ({\it Herschel}) data to further separate starless and protostellar cores
and derive core dust temperatures. Core masses have been computed assuming a dust opacity of
$k_{\nu} (\nu = 345 {\rm GHz}) = 0.86$ g$^{-1}$ cm$^{2}$ (for consistency with Massi et al.\ \cite{massi07}
and Olmi et al.\ \cite{holmes}), a dust spectral index of $\beta = 2$, and a gas-to-dust ratio of 100.
Furthermore, we used Gaia DR2 data to test the distance to Vela C, obtaining a value of 950 pc, but
we adopted 700 pc for consistency with previous works, pending further analyses
based on Gaia DR2 data. This does not affect our conclusions
and in any case scaling factors to shift our results to 950 pc are given in the Appendix. 
Our main results are summarised as follows. 
 
   \begin{enumerate}
      \item 
       A comparison between the emission at 345 GHz and that at 250 GHz obtained from SIMBA
       observations in two sub-regions of Vela C, namely IRS~31 and RCW~36, shows that even
       where intense stellar feedback is expected, the bulk of submm emission is dominated
       by dust thermal emission at 10--30 K with an emissivity index in the range 
       $\beta \sim 1 - 2$.

      \item  
         The dense cores were classified into protostellar (233 out of 549)
         and starless (316 out of 549) essentially on the basis of  
         their association with red WISE/MSX point sources. We estimated that
         this allows the detection of protostars with luminosities down to 
         $L_{\rm bol} \sim 2-3$ $L_{\sun}$. 
        Most of
         the starless submm cores lie in a region of the mass--size diagram where gravity 
          dominates over
         thermal support, and are therefore likely to be prestellar in nature.

      \item 
         A dust temperature was assigned to the dense submm cores 
         by positionally associating them with 
         {\it Herschel} compact sources. Unfortunately, this 
         only yielded dust temperatures
         for a subsample of 197 submm cores.


      \item 
        The dust temperature distribution is different for starless and
        protostellar cores, with starless cores constrained in the range of 9--12 K and
        protostellar cores peaking at 11 K with a tail at higher temperatures.

      \item 
        The core masses have been derived both for the subsample of cores associated
        with {\it Herschel} sources using the FIR-derived temperatures (single-source
         temperature), and for the whole
        sample of submm cores by assuming a constant temperature $T_{dust} = 12.5$ K. The
        latter approach allows for a more robust statistical analysis. Starless cores have
        median and average masses of $\sim 4-7$ $M_{\sun}$ and $\sim 6-10$ $M_{\sun}$, respectively,
        whereas protostellar cores have median and average masses of $\sim 8$ $M_{\sun}$ and $\sim 15$ 
        $M_{\sun}$, respectively.  
         Protostellar cores exhibit a size distribution which is less compact
        than that of prestellar cores. 

       \item 
        A core mass function was derived and its high-mass end was fit by a ${\rm d}N/{\rm d}M \sim M^{\alpha}$
        function. The power-law index $\alpha$ was also derived using the ML 
        estimator, which is less prone to statistical bias. 
        Assuming $T_{\rm dust} = 12.5$ K, we estimate a mass completeness limit
        of $\sim 3.7$ $M_{\sun}$. The effects of dust opacity, distance, and calibration
        errors and of the temperature are discussed in the text, leading to
        various determinations of 
        $\alpha$. These are mostly consistent with a Salpeter IMF ($\alpha \sim -2.35$).
        There are some indications of the protostellar CMF being slightly flatter than the
        starless CMF. This is probably due to temperature effects and/or source confusion
        in crowded areas. The ML estimator
        suggests that the starless CMF is actually slightly steeper than
        a Salpeter IMF ($\alpha = -2.8 \pm 0.2$), consistent with more recent
        IMF determinations. 

      \item 
        A comparison with theoretical works suggests that turbulence
        and magnetic field are likely to be instrumental in explaining core formation
        in Vela C.

      \item 
        The dense cores tend to cluster into groups with distances in the range 1--6 pc, 
        with hints of partial segregation of starless and protostellar cores. This probably reflects
        fragmentation of magnetised and/or turbulent clumps with density $\sim 10^{3}$ cm$^{-3}$.

      \item 
        There are strong indications of different levels of star formation activity in different
        parts of Vela C. The southern part of the cloud hosts the lowest fraction of
        protostellar cores ($\sim 35$ \% of the submm dense cores), whereas the central part
        hosts the largest fraction ($\sim 52$ \%). 
        The most massive cores are found in the centre,
        where massive star formation is in progress in RCW~36. However, line-of-sight
        effects with core merging are expected to be more frequent in the centre of the cloud. 

       \item We propose that dense cores are distributed in filaments roughly contained in sheets,
        which are seen almost face on in the case of the nests and edge-on in the case of the
        ridges. Massive star formation in the Centre-Ridge may have been triggered by converging flows.

   \end{enumerate}

\begin{acknowledgements}
This paper is based on data acquired with the Atacama Pathfinder EXperiment (APEX).
APEX
is a collaboration between the Max Planck Institute for Radioastronomy, the European
Southern Observatory, and the Onsala Space Observatory.
The authors acknowledge the help and the support from the APEX staff and the
observers for their contribution in obtaining this dataset.
This publication makes use of data products from the Wide-field Infrared Survey Explorer, 
which is a joint project of the University of California, Los Angeles, and the Jet Propulsion 
Laboratory/California Institute of Technology, funded by the National Aeronautics and 
Space Administration.
This work is also based in part on observations made with the Spitzer Space Telescope, 
which is operated by the Jet Propulsion Laboratory, California Institute of Technology 
under a contract with NASA.
This research has also made use of the NASA/ IPAC Infrared Science Archive, which is 
operated by the Jet Propulsion Laboratory, California Institute of Technology, under contract 
with the National Aeronautics and Space Administration.
This work has made use of data from the European Space Agency (ESA) mission
{\it Gaia} (\url{https://www.cosmos.esa.int/gaia}), processed by the {\it Gaia}
Data Processing and Analysis Consortium (DPAC,
\url{https://www.cosmos.esa.int/web/gaia/dpac/consortium}). Funding for the DPAC
has been provided by national institutions, in particular the institutions
participating in the {\it Gaia} Multilateral Agreement.
This research makes also use of data products from the Midcourse Space Experiment. 
Processing of the data was funded by the Ballistic Missile Defense Organization with additional 
support from NASA Office of Space Science. 
T.\ Cs.\ acknowledges support from the \emph{Deut\-sche For\-schungs\-ge\-mein\-schaft, DFG\/}  
via the SPP (priority programme) 1573 `Physics of the ISM'. 
\end{acknowledgements}


\Online

\begin{appendix} 
\section{Distance to Vela C}
\label{sec:dist}

\begin{figure}
\centering
\includegraphics[angle=-90,width=8cm]{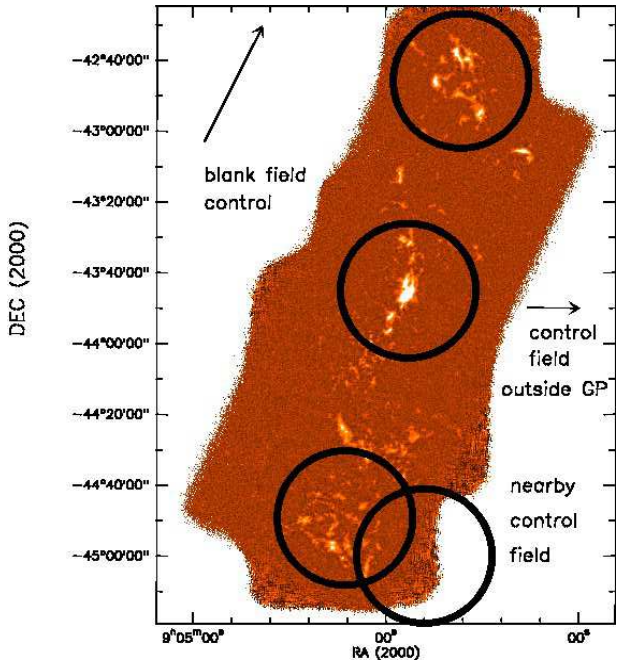}
\caption{Selected fields overlaid on the LABOCA map of Vela C}
\label{mapp:ina}
\end{figure}

\begin{figure*}
\centering
\includegraphics[width=16cm]{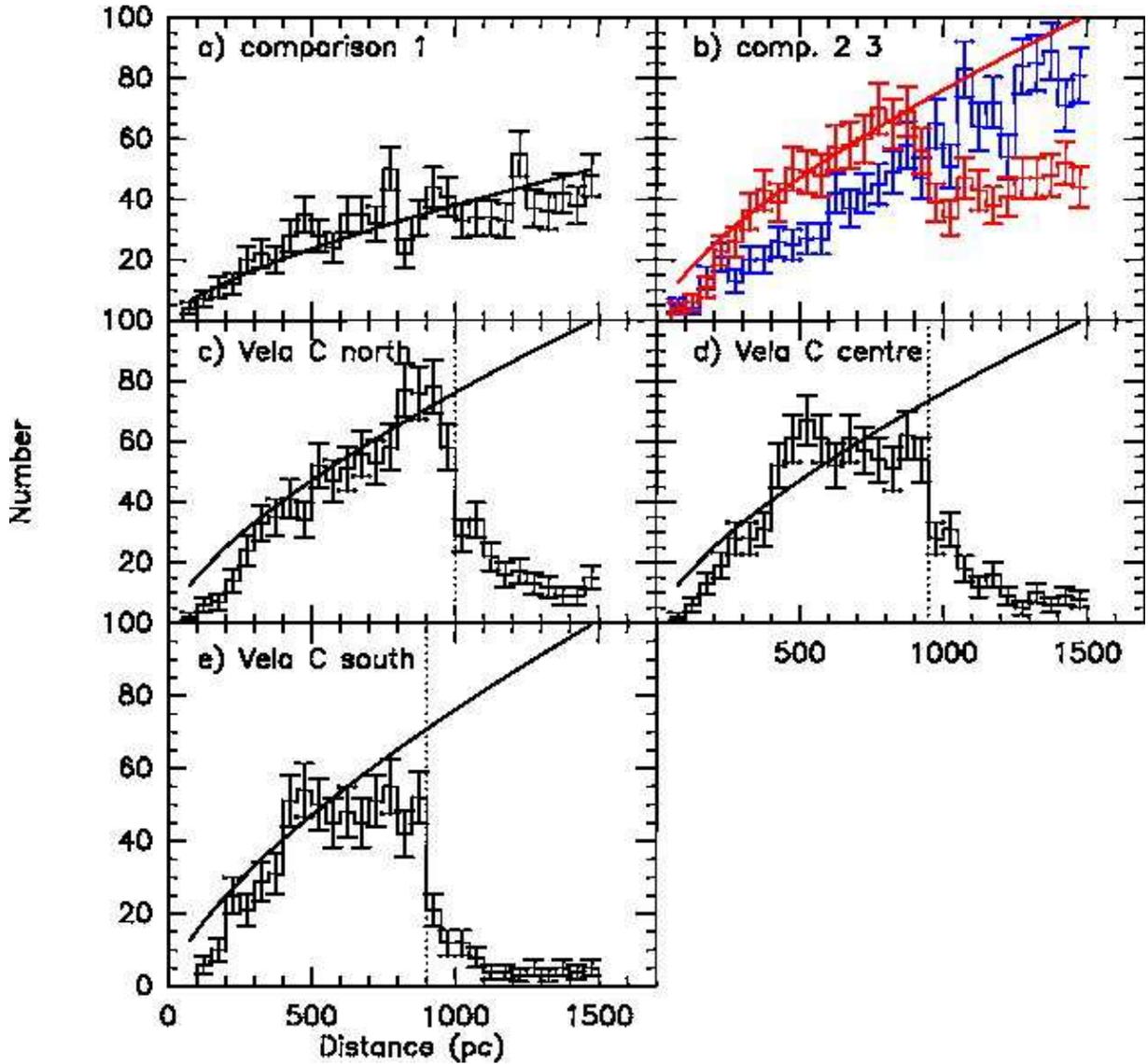}
\caption{Distance of stars in the line of sight towards three Vela C fields
and three control fields (see text), from Gaia DR2. The number of stars is
displayed in bins of 50 pc in width, error bars are computed according to Poissonian
statistics. {\bf a} Farther control field
outside the Galactic Plane, {\bf b} Nearer control field outside the Galactic
Plane (red) and control field located at $l=240\degr, b=0\degr$ (blue),
{\bf c} Field centred on sub-field North, {\bf d} Field centred on RCW36, and
{\bf e} Field centred on the South-Nest. The full line is a $N \sim d^{0.7}$ fit
to the data in {\bf a}, multiplied by two in {\bf b} to {\bf e}. The vertical dotted
lines mark the sudden decrease in the number of stars, which is likely to coincide
with the location of the molecular gas.
\label{appfig}}
\end{figure*}

Stellar parallaxes from Gaia DR2 (Gaia collaboration 2016, 2018) 
can be used to test the distance to Vela C. In this 
respect, we are interested in checking that Vela C lies in the range of values derived
by Liseau et al.\ \cite{liseau} rather than revise it.  Thus,
we retrieved stellar parallaxes from the Gaia archive\footnote{https://archives.esac.esa.int/gaia} 
in a few circular fields of  $20\arcmin$ 
in radius, so that they roughly span the width of the molecular gas distribution
when centred inside Vela C. We selected sources with a parallax-to-error ratio
$\ge 5$ to avoid the bias effects discussed in Bailer-Jones \cite{bailer}. We selected 
three fields inside Vela C (see Fig.~\ref{mapp:ina}), 
one centred on sub-field North, one centred on RCW36, and
one centred on the South-Nest. In addition, we selected three control fields, one
located $\sim 0.3$ hr west of RCW36 (i.e. outside the Galactic plane), 
one located $\sim 0.003$ hr 
west of the South-Nest, and one located at $l=240\degr, b=0\degr$. In particular,
the latter lies inside the "blank field", which is an area in the Galactic Plane adjacent
to the VMR but devoid of molecular emission (Lorenzetti et al. \cite{loren}). 
All parallaxes were
converted into distances and counted inside bins of 50 pc in width. The data are shown in
Fig.~\ref{appfig}. The farther field outside the Galactic plane (Fig.~\ref{appfig}a)
clearly shows that the number $N$ of stars increases steadily with distance $d$. The data have 
been locally fitted by $N \sim d^{0.7}$ (solid line in figure). As a first approximation, the same relation 
should hold
towards Vela C provided it is multiplied by a  constant $> 1$ (to account for the
increase in $N$ towards the Galactic Plane). In fact, when multiplied by
2 it roughly fits 
the nearest control field (histogram in Fig.~\ref{appfig}b, red line)
up to $\sim 1000$ pc, where the number of stars suddenly decreases. This should
mark absorption from molecular gas associated with Vela C,  as the control field
located in the blank field (blue histogram in Fig.~\ref{appfig}b), though not fitted by 
the $N - d^{0.7}$ relation, nevertheless exhibits a steady increase in $N$. One has
to keep in mind that, unlike the blank field, the line of sight to Vela C is roughly 
tangential to the local arm (e.g. Hou \& Han \cite{houehan}), 
which is not taken into account in our scaled $N - d$ relation.

The three fields inside Vela C (Fig.~\ref{appfig}c, d, e) clearly display an abrupt decrease in the
number of stars between $900$ and $1000$ pc, like that in the nearest control field. By comparing
with the farthest control field, we believe that this is caused by the shielding 
effect of the molecular gas. This means that stars behind the gas are so extincted that the
number of stars bright enough to allow parallax determination decreases strongly.
The Gaia data suggest a distance of $\sim 950 \pm 50$ pc.
Interestingly, a shift from 900 to 1000 pc is evident from south to north, suggesting an
extent of $\sim 100$ pc along the line of sight, roughly comparable to the projected length
of Vela C ($\sim 36$ pc).

The fields towards RCW36 and the South-Nest also exhibit a plateau between 500 and 900 pc which
is barely consistent with Poisson fluctuations. As we are observing along the
local arm, a superposition of local clouds and stellar groups is expected. To disentangle
the different components and confirm the distance to Vela C, a study of stellar colours as
a function of parallaxes is necessary, which is beyond the scope of this paper. 
Pending a careful analysis of the stellar populations in the line of sight towards 
Vela C, and more accurate
parallax determinations from the next Gaia releases, with the data at our disposal we can exclude
the nearer distance interval derived by Liseau et al. (1992) and conclude that the region
is located between 700 and 1000 pc, with $\sim 950 \pm 50$ pc as a likely value.

We note that the updated distance would not significantly change the results
discussed in the paper. The
slopes derived through linear fits would not change. The cores would be slightly larger and
more massive, but this is not a problem, and the sensitivity to protostellar
cores only slightly worse. The differences are listed in Table~\ref{dist1:dist2}. 
From Table~\ref{dist1:dist2}, it is clear that core masses in Fig.~\ref {mass:size:fig}
are shifted vertically more than the loci of Bonnor-Ebert critical half-masses, so
the issue of gravitational boundedness is reinforced if the updated distance is assumed. 

%
%
\begin{table}
         \caption{Effects of distance on the derived parameters.} 
\label{dist1:dist2}      
\footnotesize
\centering                          
\begin{tabular}{c | c | c |}        
Parameter & at 700 pc & at 950 pc \\
\hline\\
CMF slope & $\alpha$ & $\alpha$ \\
\hline\\
log(size)-log(mass) slope & $x$ & $x$ \\
\hline\\
Sizes & $D$ & $1.36 \times D$ \\
and linear resolution & & \\
\hline\\
Masses &  $M$ & $1.84 \times M$ \\
and mass completeness limit & & \\
\hline\\
70 $\mu$m sensitivity & $0.28$ $L_{\sun}$ & $0.5$ $L_{\sun}$ \\
to protostars & & \\
\hline\\
WISE sensitivity & $1.7 - 2.8$ $L_{\sun}$ & $3.1 - 5.0$ $L_{\sun}$ \\
to protostars & $0.4 - 0.5$ $M_{\sun}$ & $0.6 - 0.7$ $M_{\sun}$  \\
\hline\\
Bonnor-Ebert critical mass & $m_{\rm BE}$ & $1.36 \times m_{\rm BE}$ \\
as a function of {\it apparent} size & & \\
\hline
\end{tabular}
\end{table}

\section{Comparison of core masses towards RCW~36}
\label{comp:app}

Minier et al.\ \cite{minier}  
found seven clumps surrounding RCW~36 by combining FIR
{\it Herschel} data and submm (450 $\mu$m) ground-based data. All these structures are
retrieved by CuTEx in our LABOCA map, but for only four we were able to assign a
dust temperature from the association with the {\it Herschel} cores of Giannini et al.\ \cite{gianni12}. 
The masses computed from the submm emission mapped with LABOCA (see Table~1) by
using these temperatures are compared with those computed by Minier et al.\ \cite{minier}
in Table~\ref{mass:mass}.  Although the masses from our data are
consistent with the ranges of values that they indicate 
for their sources 5 and 7, sources 1 and 3 are much less massive
than we find.
They adopted temperatures of 17--25 K to derive the ranges of masses, which are
consistent with temperatures of 20--21 K found from the
{\it Herschel} data. In addition, their
adopted opacity would result in $1.2$ cm$^{2}$ g$^{-1}$ at 345 GHz, as opposed
to our adopted $0.86$ cm$^{2}$ g$^{-1}$, that is only 30 \% more. 
From Table~2 of Minier et al.\ \cite{minier}, the highest masses are obtained
by measuring the flux at 350 $\mu$m in a circular aperture $25\arcsec$ in diameter.
This is less than the size of the corresponding submm cores (see Table~1). On the other
hand, they find lower masses by combining fluxes at 160 and 450 $\mu$m with a better
spatial resolution ($12\arcsec$), which suggests a higher degree of source confusion
at lower spatial resolutions. In fact, in Sect.~\ref{sfavc} we propose that
source confusion leading to the merging of unresolved sources may yield the most massive
cores in the centre region. This inconsistency
highlights real differences arising from combining data at different wavelengths with
different sensitivities and spatial resolutions. 

%
%
\begin{table}
         \caption{Comparison of the masses of cores towards RCW~36 
           derived by Minier et al.\ 
           \protect\cite{minier} and from our LABOCA data (columns 3--4).}             
\label{mass:mass}      
\footnotesize
\centering                          
\begin{tabular}{c | c | c | c }        
\hline
Source ID & Mass & Source ID & Mass \\
in Minier et al.\ \protect\cite{minier} & & in Table~1  \\
        & ($M_{\sun}$)    &           &   ($M_{\sun}$)  \\
\hline
1 & 17--72 & 453 & 128 \\
3 & 4--25 & 449 & 115 \\
5 & 10--25 & 446 &  9 \\
7 & 3--19 & 457 & 17 \\
\hline
\end{tabular}
\end{table}

\end{appendix}


\end{document}